\documentclass{aa}
\usepackage{psfig}
\usepackage{natbib}
\newcommand\cf{{cf.,~}}
\newcommand\eg{{e.g.,~}}

\newcommand\ie{{i.e.,~}}
\def\spose#1{\hbox to 0pt{#1\hss}}
\newcommand\simlt{\mathrel{\spose{\lower 3pt\hbox{$\mathchar"218$}}
     \raise 2.0pt\hbox{$\mathchar"13C$}}}
\newcommand\simgt{\mathrel{\spose{\lower 
3pt\hbox{$\mathchar"218$}}
     \raise 2.0pt\hbox{$\mathchar"13E$}}}

\begin{document}

\title{Determination of the gas-to-dust ratio in nearby dense clouds using X-ray absorption measurements}

\subtitle{}

\author{M.H.  Vuong \inst{1} \and T.  Montmerle \inst{1,2} \and N.  
Grosso \inst{2} \and E.D.  Feigelson \inst{1,3} \and L. Verstraete \inst{4} \and H.  Ozawa \inst{1}}

\offprints{M\~y H\`a Vuong\\
e-mail: vuong@discovery.saclay.cea.fr}

\institute{Service d'Astrophysique, CEA Saclay, 91191 Gif-sur-Yvette, France \and Laboratoire d'Astrophysique de Grenoble, Universit\'e Joseph Fourier, BP 53, 38041 Grenoble Cedex 9, France \and Department of Astronomy \& Astrophysics, 525 Davey Laboratory, Pennsylvania State University, University Park, PA 16802, USA \and Institut d'Astrophysique Spatiale, B\^at. 121, Universit\'e de Paris XI, 91405 Orsay Cedex, France}

\date{Received 2003 March 17 /Accepted 2003 June 20}
\authorrunning{Vuong et al.}
\titlerunning{Determination of the gas-to-dust ratio in dense clouds 
using X-ray absorption measurements}

\abstract{

We present a comparison of the gas and dust properties of the dense interstellar matter in six nearby star$-$forming regions ($d<$500 pc): $\rho$ Oph, Cha~I, R~CrA, IC~348, NGC~1333, and Orion. We measure from {\it Chandra} and {\it XMM$-$Newton} observations the X-ray absorption toward pre-main sequence stars (PMS) without accretion disks (i.e., Class~III sources) to obtain the total hydrogen column density $N_\mathrm{H,X}$.  For these sources we take from the literature the corresponding dust extinction in the near$-$infrared, $A_\mathrm{J}$, or when unavailable we derive it from SED fitting using the available {\it DENIS}, {\it 2MASS}, {\it ISOCAM} and other data. We then compare $N_\mathrm{H,X}$ and $A_\mathrm{J}$ for each object, up to unprecedently high extinction.  For the $\rho$ Oph dark cloud with a relatively large sample of 20 bona-fide Class~III sources, we probe the extinction up to $A_\mathrm{J} \simlt 14$ ($A_\mathrm{V} \simlt 45$), and find a best-fit linear relation $N_\mathrm{H,X}/A_\mathrm{J}$ = 5.6 ($\pm$ 0.4) $\times$ 10$^{21}$ cm$^{-2}$ mag$^{-1}$, adopting standard ISM abundances. The other regions reveal a large dispersion in the $N_\mathrm{H,X}$/$A_\mathrm{J}$ ratio for each source but for lack of adequate IR data these studies remain limited to moderate extinctions ($A_\mathrm{J} \simlt 1.5$ or $A_\mathrm{V} \simlt 5$). For $\rho$ Oph, the $N_\mathrm{H,X}/A_\mathrm{J}$ ratio is significantly lower ($\simgt 2\sigma$) than the galactic value, derived using the standard extinction curve ($R_\mathrm{V} = 3.1$).
This result is consistent with the recent downwards revision of the metallicity of the Sun and stars in the solar vicinity. We find that the $\rho$ Oph dense cloud has the same metallicity than the {\it local} ISM when assuming that the galactic gas-to-dust ratio remains unchanged. The difference between galactic and local values of the gas-to-dust ratio can thus be attributed entirely to a difference in metallicity.

}

\maketitle

%

\section{Introduction}\label{introduction}

\begin{table*}
\caption[]{Methods for determining $N_\mathrm{H}$/$A_\mathrm{V}$ ratios.}\label{NH_AV_ratio}
\small
\begin{tabular}{cccccc}
\hline
\hline
Method & $A_\mathrm{V}$ max & Medium & Length scale & Ref.&\\
\hline
B stars (UV)     & $\sim$2 mag & diffuse local & $\sim$ few 100\,pc & \citet{BOH78} &\\
B stars (UV)     & $\sim$2 mag & diffuse local & $\sim$pc & \citet{WHI81} &\\
SNR (X-rays)     & $\sim$30 mag & diffuse galactic & $\sim$kpc & \citet{GOR75} &\\
SNR (X-rays)     & $\sim$30 mag & diffuse galactic & $\sim$kpc & \citet{RYT75} &\\
SNR + compact objects (X-rays)$^*$  & $\sim$20 mag & diffuse galactic & $\sim$kpc & \citet{PRE95}&\\
{\it Diskless T Tauri stars (X-rays)$^*$} & {\it $\sim$50 mag} & {\it dense clouds} & $\sim$ {\it sub$-$pc} & {\it This work} &\\
\hline
\end{tabular}

$^*$ X-ray spectroscopy with imaging X-ray satellites
\end{table*}

One of the key parameters to study galactic evolution, star formation, and other fundamental problems in astrophysics is the so-called ``gas-to-dust ratio''. The only way to measure this ratio is to start from the ``gas-to-extinction'' ratio, expressed in terms of the ratio of the gas (i.e., hydrogen in all forms) column density $N_\mathrm{H}$ toward some sources, to the dust extinction in front of the same source, usually expressed in near-IR or optical magnitudes $A_\mathrm{J}$ or $A_\mathrm{V}$. The dust extinction is generally derived from data covering a limited wavelength range in the UV to near-IR bands.  The actual ratio by mass ($M_\mathrm{gas}/M_\mathrm{dust}$) can then be obtained from the $N_\mathrm{H}/A_\mathrm{V}$ ratio, at the cost of some knowledge of the physical state of the gas to derive $M_\mathrm{gas}$ from $N_\mathrm{H}$, and some modeling of dust grains to go from $A_\mathrm{V}$ or $A_\mathrm{J}$ to $M_\mathrm{dust}$. Although $M_\mathrm{gas}/M_\mathrm{dust}$ {\it a priori} depends both on the length and direction of the line-of-sight, the well-known remarkable result is that this ratio has almost the same value (with some dispersion) throughout the Galaxy, with $M_\mathrm{gas}/M_\mathrm{dust} \simeq 100$.

Perhaps the most widely known method to determine the $N_\mathrm{H}/A_\mathrm{V}$ ratio is to use the Ly$\alpha$ absorption to measure the H~{\small I} column density in front of hot, UV emitting OB stars, and use their spectral types and $B-V$ color excess to derive the extinction. Such measurements have been performed in the past using the {\it Copernicus} \citep[]{BOH78} and {\it IUE} satellites \citep[]{SHU85,DIP94}. {\it Copernicus} also provided a complementary measure of the H$_2$ column density from the rotational and vibrational levels of hydrogen in the UV. However, the method is limited to low extinctions since UV photons can only penetrate up to $A_\mathrm{V} \simlt 2-3$. 

In this paper, we measure hydrogen column densities using X-ray absorption toward young stars. The method is conceptually similar to the classical method based on Ly$\alpha$ absorption in the UV, with two major improvements~:

\begin{enumerate}

\item X-ray absorption measures the {\it total} hydrogen column density (solid state, molecular, atomic and ionized), contrary to Ly$\alpha$ absorption which measures only the {\it atomic} hydrogen column density.  

\item X-rays can probe {\it much higher extinctions} (with the single assumption of a metallicity ratio) than the UV range: several tens of magnitudes or more (depending on the X-ray energy range available), as opposed to only a few magnitudes for the UV, which is ideal for the study of dense material.

\end{enumerate}

More precisely, X-rays are absorbed along the line of sight by photoelectric effect on heavy atoms \citep[]{MOR83,WIL00}.  The X-ray extinction cross section varies as E$_X^{-2.5}$, hence rapidly decreases at high X-ray energies \citep[\eg][]{RYT96}.  In dense regions such as the cores of molecular clouds or the immediate vicinity of young stellar objects (YSO), self-shielding effect may become important giving a reduced contribution to the X-ray absorption.  Except for this effect which concerns only the relatively minor K-shell edges of refractory elements, partially ionized, atomic, molecular and solid phases contribute equally to the X-ray absorption. In other words, the X-ray absorption is dominated not by the H atoms themselves but by metals, mainly C, N, and O, and is almost insensitive to the presence of grains along the line-of-sight \citep[\eg][]{MOR83}.

The method of using X-rays to determine the total hydrogen column density is not new. It has been applied as early as 1975 from X-ray spectra obtained using scanning X-ray collimators aboard rockets or satellites. \citet{GOR75} and \citet{RYT75} derived relations between interstellar X-ray absorption and optical extinction, from the correlation of measurements of supernova remnants (SNRs) and the optical $E(B-V)$ color excess, up to 
$A_\mathrm{V} \sim 30$.  More recently, \citet{PRE95} used {\it ROSAT} images to study X-ray halos produced by dust scattering around bright compact point sources and supernova remnants, and determined with this technique dust extinctions up to $A_\mathrm{V} \simlt 20$. All these studies have assumed solar abundances for the ISM absorption.

However, all the above methods probe mostly the low-density, {\it diffuse} interstellar medium.  High extinctions are reached in these X-ray studies only because the line-of-sight is long, typically on the kpc scale, so that many local effects are averaged out, not only in depth but also in projection on the plane of the sky (in the case of SNRs, which are extended X-ray sources).

\begin{table*}
\caption[]{{\it Chandra} data used in this work.}\label{observations}
\begin{tabular}{cccccc}
\hline
\hline
Region name    & Position (J2000)  & Obs ID & Observation date  & Exposure time & Ref.     \\
\hline
$\rho$ Oph cores B+F & 16$^{\mathrm h}$27$^{\mathrm m}$17$^{\mathrm s}$ $-$24$\degr$34$\arcmin$39$\arcsec$ & 635 & 2000-04-13 & 100ks & \citet{IMA01a, IMA01b, IMA02} \\ 
$\rho$ Oph core A    & 16$^{\mathrm h}$26$^{\mathrm m}$34$^{\mathrm s}$ $-$24$\degr$23$\arcmin$28$\arcsec$ & 637 & 2000-05-15 & 100ks &  \citet{DAN00} \\
$\rho$ Oph mosaic    & 16$^{\mathrm h}$25$^{\mathrm m}$32$^{\mathrm s}$ $-$24$\degr$25$\arcmin$03$\arcsec$ & 618 & 2000-06-22 & 5ks  & {\it This work} \\ 
$\rho$ Oph mosaic   & 16$^{\mathrm h}$28$^{\mathrm m}$26$^{\mathrm s}$ $-$24$\degr$17$\arcmin$33$\arcsec$ & 623 & 2000-02-08 & 5ks &  {\it This work} \\ 
Cha~I                   & 11$^{\mathrm h}$10$^{\mathrm m}$12$^{\mathrm s}$ $-$76$\degr$34$\arcmin$30$\arcsec$ & 1867 & 2001-07-02 & 70ks &  {\it This work} \\ 
R CrA                     & 19$^{\mathrm h}$01$^{\mathrm m}$50$^{\mathrm s}$ $-$36$\degr$57$\arcmin$30$\arcsec$ & 19 & 2000-10-07 & 20ks &  {\it This work} \\ 
IC\,348                  & 03$^{\mathrm h}$44$^{\mathrm m}$30$^{\mathrm s}$ $+$32$\degr$08$\arcmin$00$\arcsec$ & 606  & 2000-09-21 & 50ks & \citet{PZ01,PZ02} \\ 
NGC\,1333                 & 03$^{\mathrm h}$29$^{\mathrm m}$05$^{\mathrm s}$ $+$31$\degr$19$\arcmin$19$\arcsec$ & 642 & 2000-07-12 & 50ks & \citet{GET02} \\ 
Orion                   & 05$^{\mathrm h}$35$^{\mathrm m}$15$^{\mathrm s}$ $-$05$\degr$23$\arcmin$20$\arcsec$ & 18 & 1999-10-12 & 47ks & \citet{FEI02} \\
Orion                   & 05$^{\mathrm h}$35$^{\mathrm m}$15$^{\mathrm s}$ $-$05$\degr$23$\arcmin$20$\arcsec$ & 1522 & 2000-04-01 & 36ks & \citet{FEI02} \\

\hline
\end{tabular}
\end{table*}

\normalsize

In the present paper, we want to address this issue for the first time in the case of nearby ($d \leq 500$ pc), {\it dense} molecular clouds, where the line-of-sight is dominated by H$_2$.  The high extinction is here the result not of a long line-of-sight, but of the high densities of molecular clouds ($n > 10^5$ cm$^{-3}$); the relevant length scale is here {\it small}, on the order of parsecs or less.  The idea is to make use of the fact that ``young stellar objects'' (YSOs) are ubiquitous X-ray emitters, and at the same time may suffer high extinction, either because they are deeply embedded in, or because they are located behind, dense material.  We show below that after having defined appropriate selection criteria, we are able to construct a reliable sample of X-ray ``candles'' allowing the derivation of $N_\mathrm{H}$ (from X-ray absorption) and $A_\mathrm{J}$ (applying standard methods) for each object, up to $A_\mathrm{J} \sim 15$ (equivalent to $A_\mathrm{V} \approx 45$), and potentially even more.  This is now possible because of the spectroscopic capabilities of the new generation of X-ray observatories ({\it Chandra}, {\it XMM-Newton}) with much higher sensitivity and angular resolution in a wide (0.5$-$10 keV) energy band.

Table \ref{NH_AV_ratio} summarizes the above mentioned approaches, for comparison with that used in the present paper.

The origin of the X-ray emission from T Tauri stars (TTS) is an enhanced solar-type activity generated by powerful magnetic reconnection flares \citep[see review by][]{FEI99}.  X-ray emission is caused by thermal bremsstrahlung and emission lines from an optically thin coronal plasma confined by magnetic loops, at temperatures $T_\mathrm{X}$ $\sim$ 10$^6 - 10^8$ K ($\sim$ 0.1$-$10 keV). The average $L_\mathrm{X}/L_\mathrm{bol}$  for T Tauri stars is $\sim$ 10$^{-4\pm1}$, or $10^2-10^4$ times higher than the active Sun, enabling their detection up to large distances. 

Obtaining $A_\mathrm{V}$ measurements for the dust component of highly obscured young stellar objects can be difficult because the $V$-band flux is often not measurable and the infrared (IR) spectrum may be dominated by local circumstellar emission. For this reason, we select a sample of X-ray emitting stars without circumstellar envelopes or disks.  In the contemporary classification scheme of pre-main sequence evolution, such stars are identified as Class~III sources in the near-IR, or `weak-lined' T Tauri stars (WTTS) in the optical \citep[\eg][]{LAD91}. The WTTS are characterized by the small equivalent width of their H$\alpha$ emission lines (EW(H$\alpha$)$<$ 10\AA) \citep[e.g.,][]{MAR98}. The spectral energy distributions (SED) of Class~III sources is consistent with a simple reddened blackbody photosphere.

In this paper, we present a study of the foreground extinction of the bona-fide Class~III stars in several nearby molecular clouds based on {\it Chandra} and {\it XMM-Newton} observations and IR data.  We describe the determination of the gas column density from the X-ray spectra (\S \ref{X-data}) and the dust extinction from the IR data (\S \ref{IR-data}) for various regions.  Next, we compare the gas column density and the IR extinction leading to $N_\mathrm{H,X}/A_\mathrm{J}$ (\S \ref{new_ratio}), comparing with the galactic value and considering the spatial dependence. \S \ref{discussion} discusses the effects of grain properties, as well as that of new stellar and ISM abundances. Conclusions and implications are presented in \S \ref{conclusion}. 

\section{Determination of the hydrogen column density from X-ray absorption}\label{X-data}

\subsection{Observational context}\label{context}

\begin{figure*}
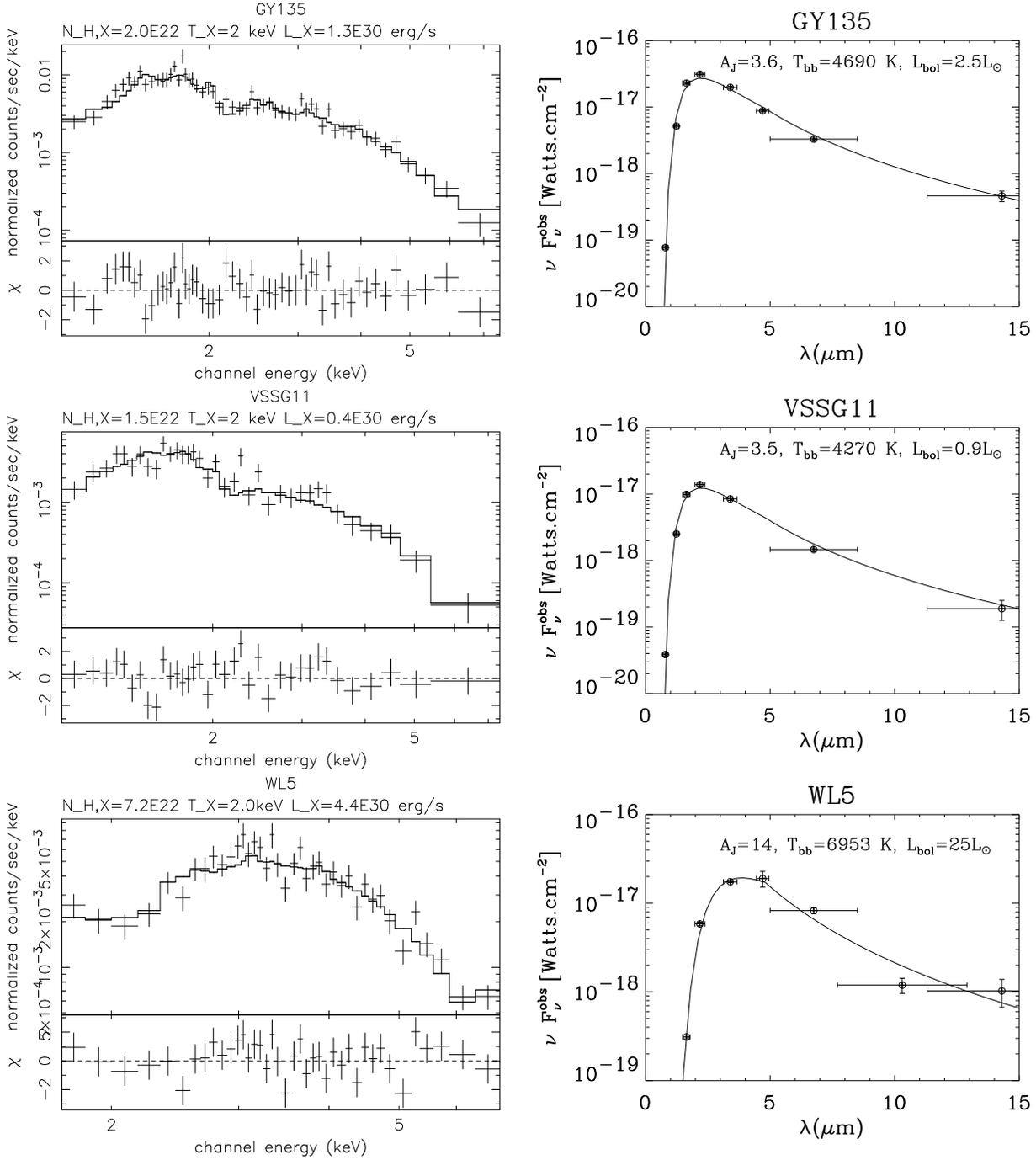

\begin{tabular}{cc}
\psfig{file=MS3731.f1a,height=6.0cm,angle=-90} &
\psfig{file=MS3731.f1b,height=6.0cm} \\
\psfig{file=MS3731.f1c,height=6.0cm,angle=-90} &
\psfig{file=MS3731.f1d,height=6.0cm} \\
\psfig{file=MS3731.f1e,height=6.0cm,angle=-90} &
\psfig{file=MS3731.f1f,height=6.0cm} \\
\end{tabular}
\caption{Sample of {\it Chandra} X-ray spectra and their corresponding IR SED fitting spectra. The left panels show X-ray spectra for three moderately absorbed $\rho$ Oph sources (GY135, VSSG11 and WL5). Their corresponding IR spectral energy distribution (SED) are in the right panels. Note the physical parameters ($N_\mathrm{H,X}$, $T_\mathrm{X}$, $L_\mathrm{X}$) derived from the X-ray spectra, using the default solar ISM abundances in XSPEC (\S 2), and the corresponding ($A_\mathrm{J}$, $T_\mathrm{bb}$, $L_\mathrm{bol}$) derived from the IR SED fitting (\S 3). See Tables \ref{Chandra_parameters} and \ref{bbfit} for the errors.}\label{X-ray spectra and SED}
\end{figure*}

The principle of deriving the hydrogen column density from the X-ray spectrum of any X-ray source is as follows.  A number of parameters can be adjusted to fit the observed spectrum with an input spectrum, which reflects the intrinsic X-ray emission mechanism of the source, in our case bremsstrahlung continuum plus emission lines from an optically thin, hot plasma (10$^6$$-$10$^8$ K).  The main fitting parameters are the absorption hydrogen column density ($N_\mathrm{H,X}$) constrained mainly by the soft energy part of the spectrum (below 1$-$2 keV) as a consequence of the energy dependence of the extinction cross section discussed above and the temperature $T_\mathrm{X}$ of the emitting plasma, which is determined by the hard part of the spectrum. It is not unusual for T Tauri stars that the spectral fit is improved by using two-temperatures plasma to describe the continuous temperature distribution of the corona. In this paper, we are only interested in deriving $N_\mathrm{H,X}$ for each source, but the accuracy of this derivation depends on the overall quality of the spectral fit, which implies a correct modeling of the plasma temperature.

We consider six nearby star forming regions observed with {\it Chandra} and {\it XMM$-$Newton}:  R CrA \citep[$\sim$ 130\,pc,][]{MAR81}, $\rho$ Oph \citep[140\,pc, see][and references therein]{GRO00}, Cha~I \citep[160\,pc,][]{WHI97}, IC\,348 \citep[310\,pc,][]{HER98}, NGC\,1333 \citep[320\,pc,][]{DEZ99} and the Orion Nebula Cluster (450\,pc).  Part of the $\rho$ Oph, NGC\,1333, IC\,348 and Orion {\it Chandra} observations have been already published by 
\citet{IMA01a}, \citet{GET02}, \citet{PZ02} and \citet{FEI02}, respectively.  These papers do quote the gas column density $N_\mathrm{H}$ but no information on the goodness-of-fit is provided, we therefore reprocessed the corresponding archived data.

To compare with previous X-ray work, we assume standard solar abundances \citep[]{AND82, GRE89, GRE98} as used by default in {\it XSPEC} for the ISM. We defer the effect of adopting more recent abundance determinations to \S \ref{discussion}.

\begin{figure*}
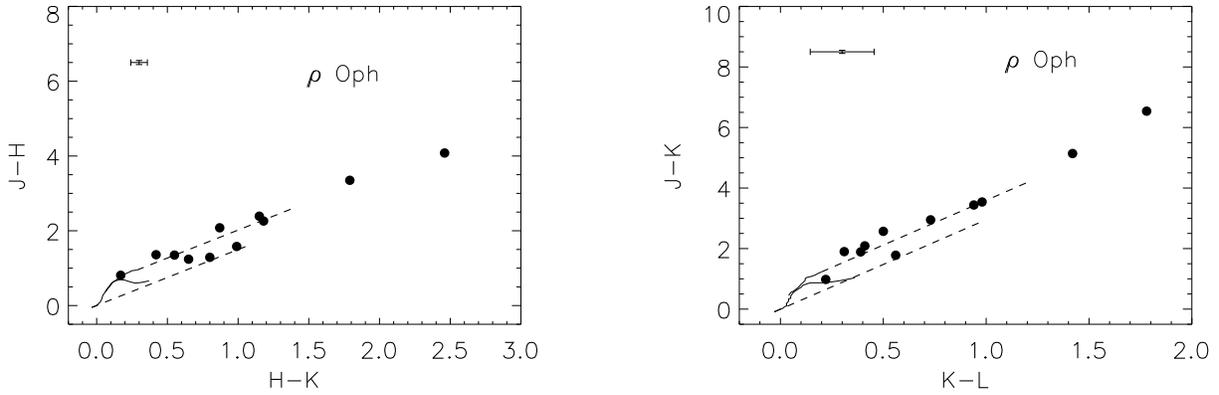
 
\begin{tabular}{cc}
\psfig{file=MS3731.f2a,width=8.5cm} &
\psfig{file=MS3731.f2b,width=8.5cm} \\
\end{tabular}
\caption{$JHK$ and $JKL$ color-color diagrams for X-ray selected ($>$ 100 counts) Class~III sources in $\rho$ Oph. The solid lines are the intrinsic colors of dwarfs and giants \citep[]{BES88}. The dashed lines extending from the extremity of the ZAMS/RGB are \citet{RL85} reddening vectors representing $A_\mathrm{J}$=5 mag ($A_\mathrm{V}$ $\simeq$ 16 mag). The typical error bar is shown in the upper left corner.} \label{JHKdiagram_oph}
\end{figure*}

\begin{figure*}
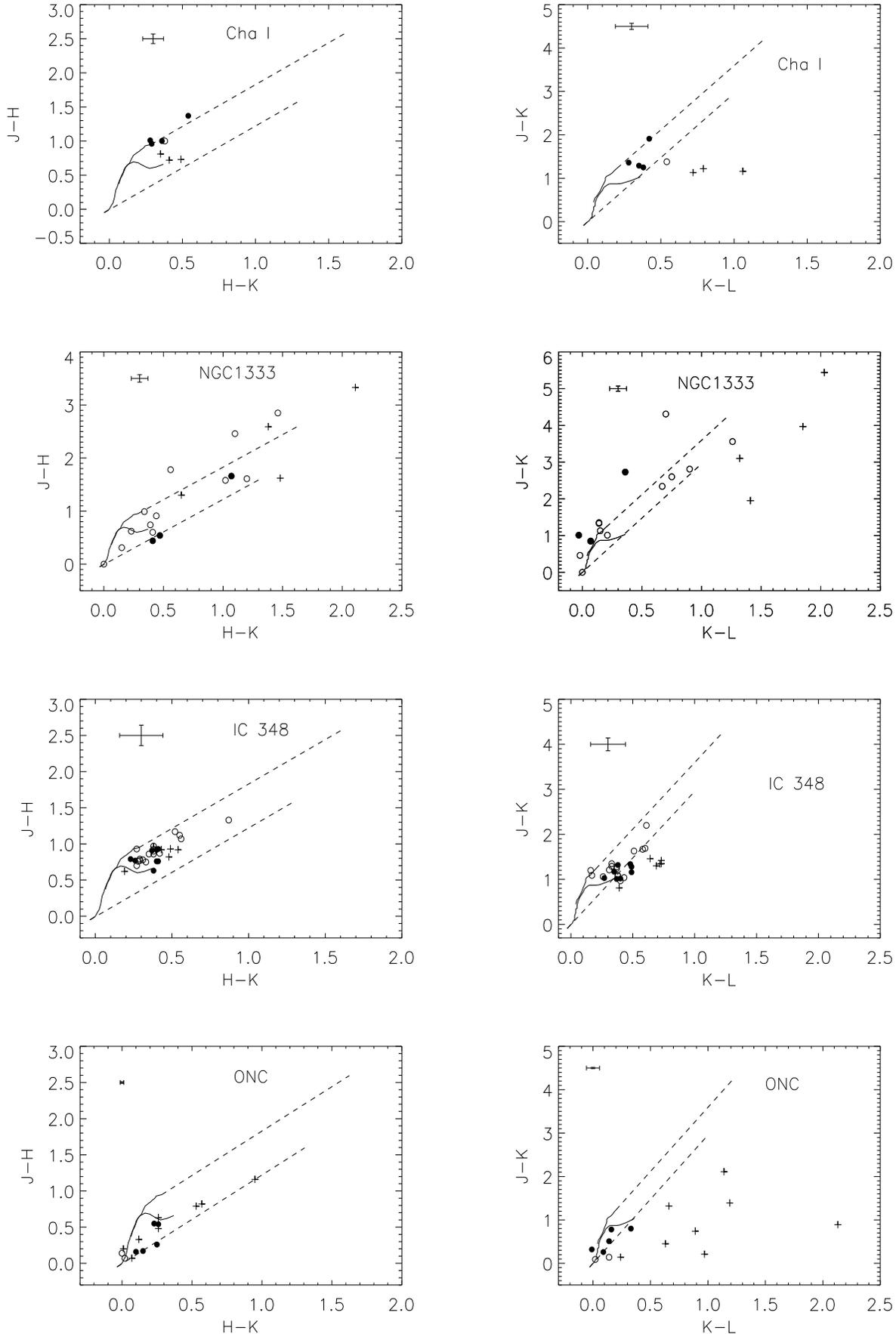
 
\begin{tabular}{cc}
\psfig{file=MS3731.f3a,width=8cm,height=6.0cm} &
\psfig{file=MS3731.f3b,width=8cm,height=6.0cm} \\
\psfig{file=MS3731.f3c,width=8cm,height=6.0cm} &
\psfig{file=MS3731.f3d,width=8cm,height=6.0cm} \\
\psfig{file=MS3731.f3e,width=8cm,height=6.0cm} &
\psfig{file=MS3731.f3f,width=8cm,height=6.0cm} \\
\psfig{file=MS3731.f3g,width=8cm,height=6.0cm} &
\psfig{file=MS3731.f3h,width=8cm,height=6.0cm} \\
\end{tabular}
\caption{$JHK$ and $JKL$ color-color diagrams for X-ray sources in Cha~I, NGC\,1333, IC\,348 and Orion Nebula. The solid lines are the unreddened ZAMS and red giant branch. The dashed lines extending from the extremity of the ZAMS/RGB are \citet{RL85} reddening vectors representing $A_\mathrm{J}$=5 mag ($A_\mathrm{V}$ $\simeq$ 16 mag). The plus signs (+) are sources representing IR L band excess, open circles ($\circ$) are faint X-ray sources ($<$ 100 counts) without IR excess and filled circles ($\bullet$) are the bright bona-fide Class~III X-ray sources with $>$ 100 counts used in this paper. The representative error bar is shown in the upper left corner.} \label{JHKdiagram_cham}
\end{figure*}

\subsection{{\it Chandra} data reduction}

All of the X-ray {\it Chandra} observations we use here are obtained with the ACIS-I array consisting of four front-side illuminated CCDs with a field of view about $17\arcmin \times 17\arcmin$ \citep[]{WEI02}.  The spectral resolution $E/\Delta E$ is $\sim$ 20. The pixel size is 0$\farcs$5 and the on-axis effective area is 600 cm$^2$ at 1.5 keV.  We start the data analysis with the Level 1 processed events provided by the pipeline processing at the {\it Chandra} X-ray Center.  We apply the techniques developed by \citet{TOW00} to correct the energy and grade of data events affected by the degradation of charge transfer inefficiency (CTI) in the energy range 0.5$-$8 keV.  We remove bad pixels, events arising from cosmic 
rays and detector noise using the ASCA grades (0,2,3,4,6) filter.  We keep events flagged as cosmic ray afterglow events for source spectral analysis to avoid loosing X-ray events on bright X-ray sources.  We use the {\it rmf} files (response matrix files) and the quantum efficiency uniformity (QEU) files provided with the CTI corrector package\footnote{http://www.astro.psu.edu/users/townsley/cti/install.html}.  These QEU files are used by {\tt mkarf } in the CIAO package\footnote{http://cxc.harvard.edu/ciao/} to generate {\it arf} files (ancillary response files) consistent with our CTI corrected events. A more detailed discussion of these data analysis methods is provided by \citet{FEI02} and \citet{GET02}.

We extract events (\ie individual photons) in the energy range 0.5$-$8\,keV for source analysis using circular apertures centered on the source position. The events are grouped into spectral bins of 15 photons using the {\it ftools} command {\tt grppha}. To have enough statistics for the spectral fitting, we keep only Class~III sources detected in X-rays with a good signal-to-noise \ie at least 100 counts.  We measure the background from a neighboring area devoid of X-ray sources. Lightcurves are studied for each source.  For the sources displaying obvious X-ray flares during the observation, we retain for the spectral analysis only the quiescent phase, since plasma abundance changes are seen to occur in YSO X-ray flares \citep[]{KAM97}, correlated with the rise of temperature.

The spectra are background subtracted and fitted with bremsstrahlung continuum models using {\it XSPEC} version 11.0.1 \citep[]{ARN96}.  The fitted models are a combination of an absorption model using the atomic cross-sections of \citet{MOR83} and an optically thin hot plasma with bremsstrahlung and emission lines using the {\tt mekal} code. The free parameters of the fit are the plasma temperature $T_\mathrm{X}$ and absorbing gas column density $N_\mathrm{H,X}$. The errors are given for 
90\% confidence intervals obtained using the command {\tt error} or {\tt steppar} in {\it XSPEC}. The coronal plasma abundance is also a parameter of the fitting (see discussion in \S 2.6). Our 1--temperature modeling is relevant if the Q$-$probability, (\ie the probability that one would observe the $\chi^2_\nu$, or a larger value, if the assumed model is true, and the best-fit model parameters are the true parameter values), is greater than 5\%. Otherwise we add another plasma temperature component to satisfy the criterion of Q$-$probability $>$ 5\%. 

Fig. \ref{X-ray spectra and SED} illustrates a sample of moderately absorbed X-ray spectra for three $\rho$ Oph sources. Table 
\ref{Chandra_parameters} presents the spectral parameters and the deduced X-ray properties of Class~III sources observed with both {\it Chandra} and {\it XMM-Newton} satellites.

\begin{table*}
\caption[]{X-ray properties of Class~III sources in $\rho$ Oph, IC\,348, R CrA, Cha~I and Orion Nebula observed with {\it Chandra} and X-ray properties of Class~III sources in $\rho$ Oph observed with {\it XMM-Newton} (using the default standard solar ISM abundances in {\it XSPEC}).}\label{Chandra_parameters}
\small
\begin{tabular}{clccccccc}
\hline
\hline
Region & Sources    & counts & $N_\mathrm{H,X}$ (10$^{22}$ cm$^{-2}$) & $\chi^2_\nu$ & $\chi^2$/d.o.f$^a$ & Q$_\mathrm{proba}$ & T(keV) & L$_\mathrm{X}$$^b$ 
(10$^{30}$erg.s$^{-1}$) \\
\hline

           & GY12   & 3730 & 3.58 (3.36$-$3.81)          & 1.07  & 117/110   & 28\%     & 2.2(2.0-2.4) & 4.91  \\
           & DoAr21 & 3097 & 1.04 (0.95$-$1.13)          & 1.14  & 133/117   & 15\%     & 2.8 (2.6-3.2) & 31.37 \\
           & S1     & 2846 & 2.43 (2.25$-$2.62)          & 1.20  & 97/81     & 11\%     & 1.14(0.93-1.37) & 3.49  \\
           &        &      &                             &       &           &          & 3.55           &       \\ 
           & GY135  & 1580 & 2.10 (1.88$-$2.22)          & 1.06  & 56.6/53   & 34\%     & 2.0 (1.8-2.3) & 1.34  \\
           & GY156  &  319 & 3.63 (3.00$-$4.36)          & 0.99  & 16.8/17   & 47\%     & 1.4 (1.1-1.8) & 0.76  \\
$\rho$ Oph & B162607 & 728 & 2.98 (2.66$-$3.31)          & 1.04  & 33/32     & 40\%     & 2.1 (1.8-2.4) & 1.12  \\
           & VSSG11 &  627 & 1.46 (1.22$-$1.75)          & 1.15  & 39/34     & 25\%     & 2.0 (1.6-2.5) & 0.40  \\
           & LFAM8  &  354 & 3.49 (3.00$-$4.15)          & 1.00  & 14/14     & 45\%     & 1.7 (1.4-2.1) & 0.57  \\
           & SR20   &  195 & 0.96 (0.45$-$1.27)          & 1.06  & 9.6/9     & 39\%     & 1.6 (1.1-3.0) & 2.29  \\

           & SR12    & 7378 & 0.07(0.04$-$0.09)          & 1.06  & 104/98    & 31\%     & 0.67(0.63-0.70) & 1.26 \\
           &             &      &                            &      &        &          & 2.1 (fixed)     &     \\ 
	   & GY289       & 388  & 3.77 (2.99$-$4.47)     & 1.01  & 16.1/16   & 44\%     & 1.9 (1.4-2.8) & 0.60  \\
           & IRS55/GY380 & 1976 & 0.96 (0.80$-$1.40)     & 1.17  & 82/70     & 15\%     & 0.7 (0.3-1.0) & 2.49  \\
           &             &      &                        &       &           &          & 2.3 (1.6-3.9) &       \\
	   & WL5         & 1322 & 7.21 (6.56$-$7.96)     & 1.24  & 47/38     & 15\%     & 2.0 (1.7-2.3) & 4.39  \\
           & WL19        & 116  & 7.56 (4.40$-$14.8)     & 0.26  &           & 85\%     & 2.90          & 0.36  \\

\hline
         & L91    & 315  & 0.41 (0.31$-$0.54) & 1.00 & 12/12   & 44\% & 2.6 (1.9-4.1)    & 0.83  \\
         & L6     & 3416 & 0.68 (0.59$-$0.81) & 1.05 & 100/96  & 35\% & 0.8 (0.7-0.9)    & 13.9  \\
         &        &      &                    &      &         &      & 2.8 (2.5-3.2)    &       \\       
IC\,348   & L154   & 136  & 0.39 (0.14$-$0.62) & 1.06 & 9.5/9   & 39\% & 1.4 (1.0-2.5)    & 0.28  \\
         & H93    & 348  & 0.42 (0.33$-$0.65) & 1.23 & 34/28   & 19\% & 0.9 (0.7-1.1)    & 0.79  \\
         & L65    & 443  & 0.41 (0.25$-$0.65) & 1.03 & 14.5/14 & 42\% & 0.7 (0.6-0.9)   & 8.82  \\
         &        &      &                    &      &         &      & 2.2 (1.7-3.2)   &       \\
	 & L58    & 361  & 0.31 (0.19$-$0.54) & 1.16 & 14/12   & 30\% & 0.8 (0.2-1.3)   & 0.83  \\
         &        &      &                    &      &         &      & 2.8 (1.7-4.4)   &       \\
         & H-IfA48& 524  & 1.13 (0.99-1.31)   & 1.26 & 24/19   & 20\% & 3.1 (2.2-5.3)   & 5.78  \\
         &        &      &                    &      &         &      & 0.6 (0.3-0.7)   &       \\
         & L36    & 532  & 0.88 (0.74-1.03)   & 1.16 & 21/18   & 29\% & 0.6 (0.5-0.8)   & 2.63  \\
         &        &      &                    &      &         &      & 1.9 (1.5-3.0)   &       \\
\hline
         & CrA PMS1  & 5141 & 0.35 (0.29$-$0.39) & 1.20 & 128/107 & 8\% & 2.7 (2.5-3.3) & 6.32 \\ 
         &           &      &                    &      &         &       & 0.4 (0.3-0.7) &    \\ 
R CrA    & CrA PMS3  & 825  & 0.37 (0.29$-$0.49) & 1.19 & 35.6/30 & 22\% & 1.5 (1.2-1.8) & 0.94 \\
         &           &      &                    &      &         &       & 0.77(0.64-0.9)&     \\     
\hline
        & CHX13a         & 1023 & 0.34 (0.26$-$0.51)  & 1.13 & 38/34 & 27\% & 0.7 (0.6-0.8)   & 0.60 \\
        &                &      &                     &      &       &      & 2.6 (2.0-4.7)   &     \\
Cha~I   & CHRX 37        & 2439 & 0.86 (0.74$-$1.00)  & 1.15 & 88/77 & 18\% & 0.22(0.17-0.24) & 16.4 \\
        &                &      &                     &      &       &      & 1.6 (1.5-1.9)   &      \\
        & GK1            & 2126 & 0.49 (0.40$-$0.57)  & 0.93 & 68/74 & 66\% & 0.8 (0.7-0.9)   & 1.17 \\
        &                &      &                     &      &       &      & 2.3 (2.0-2.5)   &      \\
        & CHRX 40        & 2954 & 0.63 (0.53$-$0.74)  & 1.14 & 92/81 & 18\% & 0.90 (0.82-0.96) & 4.62 \\
        &                &      &                     &      &       &      & 2.6 (2.1-3.5)    &     \\
        &                &      &                     &      &       &      & 0.23 (0.20-0.24) &      \\  
\hline
        & JW45           & 927  & 0.18 (0.13$-$0.25)  & 1.02 & 34.7/34 & 44\% & 0.75(0.70-0.84) & 10  \\
        &                &      &                     &      &         &      & 2.1 (1.8-2.5)   &     \\
        & JW 197         & 239  & 0.19 (0.10$-$0.42)  & 1.10 & 8.8/8   & 38\% & 1.2 (0.9-1.4)   & 1.0 \\
ONC     & JW260          & 1724 & 0.71 (0.66$-$0.89)  & 1.25 & 77/62   & 9\%  & 2.1           & 170 \\
        &                &      &                     &      &         &      & 0.6 (0.5-0.7)   &     \\ 
        & P1889          & 477  & 0.12 (0.05$-$0.19)  & 1.18 & 20/17   & 27\%   & 0.61 (0.55-0.67)  & 1.7  \\
        & P2074          & 280  & 0.68 (0.53$-$0.83)  & 1.00 & 10/10   & 76\%   & 0.25 (0.2-0.31)   & 11.9 \\
\hline
           & WL19       &  152 & 14.7(7.70-23.52)   & 1.20 & 12/10   & 24\% & 1.62(0.98-3.74) & 1.52 \\
           & GY380      & 3190 & 1.04 (0.94$-$1.15) & 1.01 & 209/207 & 43\% & 0.66-2.52       & 4.66 \\
           & GY112      & 504  & 0.41 (0.33$-$0.52) & 1.02 & 59/58   & 43\% & 0.88(0.77-1.00) & 0.53 \\
$\rho$ Oph & GY193      & 619  & 1.16 (1.04$-$1.29) & 0.94 & 43.2/46 & 59\% & 1.25(1.15-1.35) & 0.85 \\
({\it XMM})& GY194      & 262  & 1.15 (0.71$-$1.49) & 1.15 & 19/17   & 30\% & 1.25(1.0-1.73)  & 0.36 \\
           & GY253      & 704  & 4.39 (3.68$-$5.10) & 1.19 & 93.8/79 & 12\% & 1.67(1.42-2.06) & 1.97 \\
           & GY410      & 228  & 1.57 (1.21$-$1.87) & 1.21 & 35/28   & 20\% & 1.91(1.57-2.34) & 0.39 \\
           & B162730    & 494  & 0.83 (0.73$-$1.04) & 1.03 & 67/65   & 40\% & 1.87(1.54-2.23) & 0.55 \\
\hline

\end{tabular}

$^a$ degrees of freedom.\\
$^b$ Absorption corrected luminosity $L_\mathrm{X}$ 
(10$^{30}$erg.s$^{-1}$) from the X-ray spectrum in the 0.5$-$10 keV range.
\end{table*}

\subsection{{\it Chandra} observations of the $\rho$ Ophiuchi dark cloud}

The $\rho$ Ophiuchi dark cloud is a nearby low-mass star formation molecular cloud which has been the subject of numerous X-ray studies.  Early observations with the {\it Einstein Observatory} satellite showed that YSOs are bright and variable X-ray emitters in the 0.2$-$4.5 keV energy band \citep[]{MON83}. \citet{KOY94} and \citet{KAM97} detected hard X-rays from T Tauri stars with {\it ASCA} in the range 0.5-10 keV. \citet{CAS95} and \citet{GRO97, GRO00} reported deep {\it ROSAT} Position Sensitive Proportional Counter (PSPC) and the follow-up {\it ROSAT} High Resolution Imager (HRI) imaging of the $\rho$ Oph cloud dense cores A and F.  

The {\it Chandra} observations were made with the ACIS$-$I CCD array in the 0.2-10 keV band.  A large area of the $\rho$ Oph region has been covered ($\sim$ 40$\arcmin$ $\times$ 40$\arcmin$).  It consists in a 100\,ks exposure observation of cores B and F, a 100\,ks exposure observation of the core A and seven 5\,ks shallow exposures forming a mosaic (see Table \ref{observations} for details). In the core B and F region, \citet{IMA01a} identified 58 X-ray sources with near-IR counterparts.  From this data set, we select 6 T Tauri stars without mid-IR excess, i.e., without accretion disks \citep[{\it ISOCAM} data,][]{BON01}.  We re-analyze the X-ray spectrum of these sources to obtain the goodness-of-fit parameters.  Five sources have enough counts for our study.  For the other fields, we identified 8 bona-fide Class~III sources in the $\rho$ Oph A field and 3 Class III sources in the mosaics using the classification of \citet{BON01}. The selected $\rho$ Oph sources have 166 to 7400 X-ray counts. 

Fig. \ref{JHKdiagram_oph} presents the color-color diagrams $J-H$ vs. $H-K$ and $J-K$ vs. $K-L$ of our resulting sample of X-ray emitting Class~III sources in $\rho$ Ophiuchi with available $L$ band photometry. These NIR sources previously classified as Class~III using $ISOCAM$ data \citep[]{BON01} are located within the area defined by the reddening vectors. Since $ISOCAM$ data are not available for the other regions, we then use similar diagrams to define bona-fide Class~III sources as those, taking into account their color uncertainties, which lie in the area between the reddening vectors extending from the extremity of the ZAMS in both diagrams (\cf Figs \ref{JHKdiagram_cham}).

\subsection{{\it Chandra} observations of other star forming clouds}

The Chamaeleon complex is one of the star formation regions nearest to the Sun \citep[between 160 and 180 pc,][]{WHI97}. It consists of three main dark clouds, designated Cha~I, II and III.  Cha~I is the most active and hence the most studied dark cloud of the Cha complex.  X-ray observations of the Cha~I region have been obtained with {\it Einstein} \citep[]{FEI89} and with {\it ROSAT} \citep[]{FEI93}.  We use a 70ks exposure archival image of Cha~I obtained with {\it Chandra} (see Table \ref{observations}).  Eight X-ray sources are previously known WTTS and have published $JHKL$ photometry \citep[]{KEN01}. To select the TTS without circumstellar disks, we used two color-color diagrams ($J-H$ vs. $H-K$ and $J-K$ vs. $K-L$) shown in Fig. \ref{JHKdiagram_cham} and apply the same criterion as for $\rho$ Oph (Fig. \ref{JHKdiagram_oph}). We then select 4 X-ray emitting Class~III sources with more than 100 counts.

Like Cha~I, the R Coronae Australis (R~CrA cloud) has also been extensively studied in X-rays: with {\it Einstein} \cite[]{WAL81, WAL97}, with {\it ASCA} \citep[]{KOY96} and with {\it ROSAT} \citep[]{NEU00}.  In the 20\,ks archival {\it Chandra} exposure, three Class III sources are identified but only two are sufficiently bright for our study.

\begin{figure*}
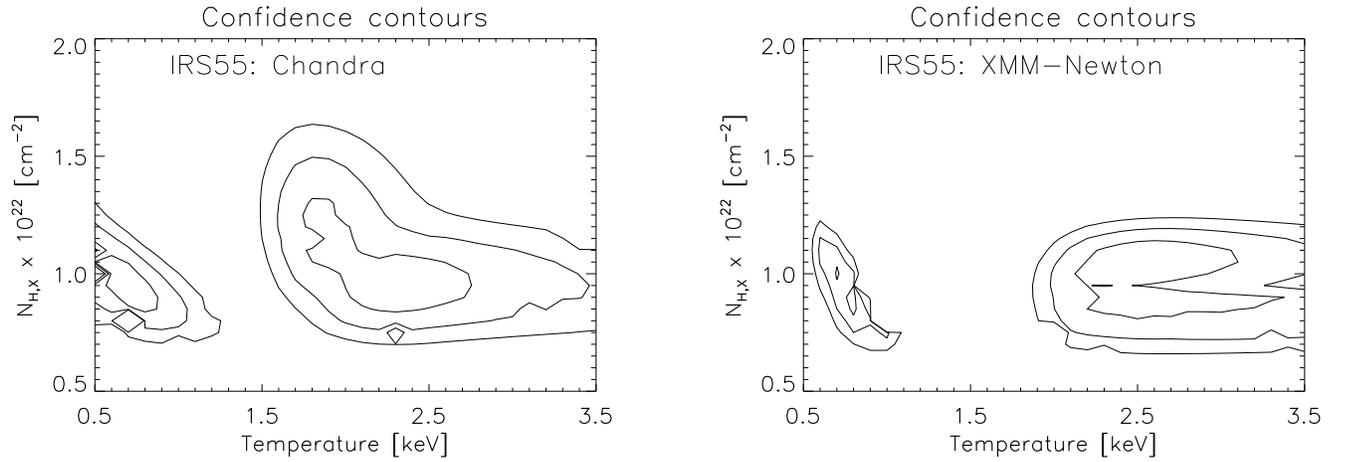
 
\begin{tabular}{cc}
\psfig{file=MS3731.f4a,width=9cm} &
\psfig{file=MS3731.f4b,width=9cm} \\
\end{tabular}
\caption{Confidence contours of $N_\mathrm{H,X}$ versus temperature spectral fit of IRS55 in the $\rho$ Oph cloud detected by both {\it Chandra} (left) and {\it XMM-Newton} (right). The contours are 1$\sigma$, 2$\sigma$ and 3$\sigma$ levels respectively. The values of $N_\mathrm{H,X}$ from these two independent observations are consistent within 1$\sigma$ confidence.}\label{IRS55_contour}
\end{figure*}

NGC\,1333, an active star forming region within the Perseus molecular complex, was previously observed and studied in X-rays by \citet{PRE97} using {\it ROSAT} and by \citet{GET02} with {\it Chandra}. In the 37.8\,ks exposure time, they detected 127 X-ray sources and identified 95 with previously known cluster members in the optical and near IR.  Spectroscopy is not available for these known cluster members. Seventeen X-ray sources are identified with $JHKL$ sources \citep[]{ASP94,ASP97} and have been plotted in the two color-color diagrams (Fig. \ref{JHKdiagram_cham}).  We find 14 sources within the area defined by the reddening vectors extending from the extremity of the ZAMS in both diagrams including the foreground B star BD+30$^{\circ}$547. These sources are symbolized as circles in both diagrams.  Three sources meet the X-ray selection criterion, but  unfortunately they fall outside of this area, so that their status is unclear \citep[see discussion in][]{ASP97}. We have not included these sources in our sample, and therefore the whole region, in spite of its interest, has to be excluded from our study for lack of adequate data.

IC\,348 is a nearby star forming region also belonging to the Perseus molecular cloud complex.  The most massive cluster member is the B5 V star BD$+$31$^{\circ}$643. {\it ROSAT} X-ray observations of IC\,348 \citep[]{PRE96} had shown that this region contains over a hundred stars. \citet{PZ01,PZ02} obtained a 50\,ks exposure image of IC\,348 with ACIS$-$I. They detected 215 X-ray sources and identified 115 of these sources with known cluster members. They used the absorption derived 
from optical/near-IR spectroscopy \citep[]{LUH98} to fit the spectra instead of deriving it from the X-ray spectra. Therefore, we have re-analyzed this dataset to derive the gas column density from our X-ray spectra. Optical/near-IR spectroscopy of this region is available \citep[]{LUH98,LUH99}.  But \citet{PZ02} noticed that many TTS selected on the basis of H$\alpha$ equivalent width EW(H$\alpha$) $<$ 10\,\AA\ have a $JHK$ infrared excess.  We first pre-select the X-ray sources 
whose counterparts have EW(H$\alpha$) $<$ 10\,\AA\ and we found 29 sources.  Next, we use $JHK$ photometry and the 3.5 $\mu$m data from \citet{HLL01} of these sources to construct the color-color diagrams (Fig. \ref{JHKdiagram_cham}). Twenty X-ray sources are located within the reddening band in both color-color diagrams which have no $L$ band excess. Eight have $>$ 100 counts required for a good fit to the X-ray spectrum.

The Orion Nebula is the most distant highly active star forming region of our sample. Observations with the {\it Einstein}, {\it ROSAT} and {\it ASCA} satellites have been carried out but could identify only a small fraction of X-ray sources of the cluster population due to sensitivity, angular resolution and bandwidth limitations.  With much higher sensitivity and angular resolution, {\it Chandra} recently detected more than 1,000 sources in the ONC identified with previously known members \citep[]{FEI02}. \citet{HIL97} studied the optical spectroscopy of $\sim$ 900 stars in this region but did not provide information on the H$\alpha$ equivalent width (to distinguish between CTTS and WTTS), due to the contamination of this line by the emission of the blister H~{\small II} region associated with the Orion Nebula.  Using $JHKLMNQ$ catalog in \citet{HIL98}, we correlate this catalog with the X-ray source list detected by \citet{FEI02} and find 16 sources which have $JHK$ and $L$ data and a good X-ray spectrum. We select seven of these sources which have no L-band excess and are located within the reddening band (Fig. \ref{JHKdiagram_cham}).  Six sources have small or zero $K-$band excess $\Delta(I-K)$ $<$ 0.3 \citep[]{HIL98}. The column density derived from the \citet{FEI02} paper is not sufficiently accurate because of the automated fitting procedure of a large number of sources.  We re-analyze the X-ray spectra of these 6 sources and find improved spectral parameters which are listed in Table \ref{Chandra_parameters}.

\subsection{{\it XMM$-$Newton} observation of the $\rho$ Ophiuchi dark cloud}

The $\rho$ Ophiuchi cloud was also observed with \it{XMM-Newton} \rm in the 0.2$-$10.0 keV band for 32 ks on 2001 February 19 centered at $\rm{R.A.}$ $=16^{\mathrm h}27^{\mathrm m}26.0^{\mathrm s}$ and $\rm{DEC}$=$-$24$\degr$40$\arcmin$48.0$\arcsec$. \it{XMM-Newton} \rm features three coaligned X-ray telescopes with 6$\arcsec$ FWHM angular resolution at 1.5 keV and with large total effective area up to 5000 cm$^{2}$ at 1 keV \citep[]{JAN01}. The European Photon Imaging Camera (EPIC) consists of two MOS (MOS\,1 and MOS\,2) and one backlighted PN, X-ray CCD arrays placed on each focal plane of the telescopes. The EPIC provides capability of imaging with a field of view of 30$\arcmin$ and spectroscopy with moderate spectral resolution of $\sim$ 60 eV at 1 keV.

The data reduction followed standard procedures using the \it{XMM-Newton} \rm Science Analysis Software (SAS version 5.3).  We ran {\tt emchain} and {\tt epchain} to obtain photon event lists.  We removed bad pixels and selected single, double, triple, and quadruple pixel events for MOS and single and double pixel events for PN using {\tt evselect}.  We extracted light curves from whole field of view and excluded the time region where the X-ray counts are extremely high, which could result from solar flare protons.

The background-subtracted X-ray spectra were obtained from 8 Class~III sources with good statistics (150-3200 counts).  The counts were extracted from the 90$\%$ enclosed energy circle. In the case other X-ray sources are found close to the Class~III sources, we adjusted the radius of the reference regions to minimize the contamination by these sources.  The background X-ray counts were extracted from annular regions around each Class III source.  The X-ray spectra were obtained from each EPIC CCD detector (MOS1, MOS2, and PN), and fit simultaneously with the same model used for \it{Chandra} \rm spectral analysis using \it{XSPEC}\rm.

We find that three sources (IRS\,55, SR\,12 and WL\,19) of our sample are in common with {\it Chandra}. This allows a test for possible systematic errors between the two satellites.

\subsection{Reliability of the X-ray derived $N_\mathrm{H,X}$ values}

We consider here several contributions to the uncertainty of the $N_\mathrm{H,X}$ values derived from the Class~III X-ray spectra: possible systematic errors associated with satellite calibrations; statistical errors from the limited source counts; possible systematic errors associated with the uncertain X-ray spectrum of the underlying star; and with uncertainties in the stellar coronal abundances.

The spectra are fitted using $\chi^2$-fit statistics with the minimization technique of Levenberg-Marquardt \citep[]{BEV92}. The errors of $N_\mathrm{H,X}$ from the model fitting are given for 90\% confidence intervals obtained using the command {\tt error} in {\it XSPEC}. Table \ref{Chandra_parameters} reports these values.

We first compare the instrumental errors of the determination of $N_\mathrm{H,X}$ between {\it Chandra} and {\it XMM-Newton} for three Class~III sources observed by both satellites.  Fig. \ref{IRS55_contour} shows the confidence contours of 
$N_\mathrm{H,X}$ and two-temperature fits for IRS55 in $\rho$ Oph. The agreement between the values of $N_\mathrm{H,X}$ from these two independent observations, which are consistent to better than 1 $\sigma$, is remarkably good, and gives us confidence in the reliability of the X-ray spectral fitting procedure.

Another factor of uncertainty may be the influence of statistics, i.e., whether the number of counts affects the determination of $N_\mathrm{H,X}$. \citet{GET02} have performed simulations with sources of known spectra using the {\tt fakeit} utility in the XSPEC package. Since our sources are physically of the same nature (T Tauri stars), their results are also valid in our case.  They found that the standard deviations of the derived column density values $\Delta$$N_\mathrm{H,X}$ ($\times$ 
10$^{22}$) range from $\sim$0.6 for sources with $\simlt$ 300 counts to $\sim$0.3 for sources with $\simgt$ 300 counts.  These values are included in the error bars given in Table \ref{Chandra_parameters}. 

We have also checked the effect of the coronal abundance on the value of $N_\mathrm{H,X}$. The coronal plasma abundance is first fixed at the best fit value. We then test with different values for each spectrum until the $\chi^2$ deteriorates beyond a Q$-$probability = 5\%. We find that changing the coronal abundance by a factor of $\sim$~2 leads to a $<$ 10\% variation in the value of $N_\mathrm{H,X}$, a range which is already comprised in the error bars.

Lastly, a systematic uncertainty in $N_\mathrm{H,X}$ could also arise due to a possible unobserved, absorbed soft component. Such a component would be {\it a priori} undetectable in case of high extinction, but could modify the spectral fit and surreptitiously affect the determination of $N_\mathrm{H,X}$. 
We have checked this by adding a fake soft component to the X-ray spectra. We have used the observed spectrum of WL5, which has the highest $N_\mathrm{H,X}$ and enough counts to study the effect of the soft component. Using {\tt fakeit} utility in the XSPEC package, we simulated the spectrum of WL 5 by adding a soft component at 0.3 keV to a hard component at 2 keV. We fitted the resulting spectrum with one-temperature model as seen with the real spectrum. We found that if there is a soft component in the spectrum of WL5 brighter than 5\% of the total flux, it would be picked up by XSPEC. This is already what we find in other sources which require a two-temperature model (like IRS\,55 just mentioned, see Fig. \ref{IRS55_contour}). For WL\,5 we conclude that no significant unseen soft component can affect $N_\mathrm{H,X}$, and {\it a fortiori} the other sources with smaller $N_\mathrm{H,X}$ are not affected.

\section{Determination of the extinction from IR photometry}\label{IR-data}

\begin{table}
\caption[]{Comparison between $A_\mathrm{J}$, and our SED blackbody temperature fitting with those from \citet{LR99} for the $\rho$ Oph sources.}\label{bbfit}
\small
\begin{tabular}{ccccc}
\hline
\hline
\multicolumn{1}{c}{} & \multicolumn{2}{c}{SED fitting$^{(a)}$} & 
\multicolumn{2}{c}{$K-$band spectroscopy$^{(b)}$} \\
\hline
Sources & $A_\mathrm{J}$ & T$_\mathrm{bb}$ & $A_\mathrm{J}$ & 
T$_\mathrm{eff}$ \\
\hline
GY12    & 5.02-5.23 & 3549-3925   & 4.6   & 3955  \\
GY135   & 3.61-3.64 & 4323-4690   & 4.4   & 4205  \\
GY156   & 5.05-5.22 & 4268-4733   & 6.1   & 3955  \\
LFAM8   & 5.51-5.81 & 3434-3873   & 6.6   & 3595  \\
VSSG11  & 3.50-3.51 & 3894-4262   & 4.0   & 3850  \\
B162607 & 4.98-5.11 & 3533-4103   & \ldots &\ldots \\
S1      & 3.57-3.81 & 17665-19935 & \ldots & 18700 \\
DoAr21  & 1.80-1.97 & 4614-4929   & 1.6    & 5080  \\
SR20    & 2.47-2.53 & 3677-3945   & 2.1    & 3955   \\
GY289   & 6.39-6.81 & 3910-4420   &\ldots  &\ldots \\
IRS55   & 1.41-1.51   & 3950-4175   &\ldots  &\ldots \\
WL5     & 12.26-13.93 & 5775-6970   &\ldots  &\ldots \\
SR12    & 0.21-0.22   & 3610-3640   & 0.20   & 3850  \\
GY112   & 0.96-1.05   & 3505-3745   &\ldots  &\ldots \\
GY193   & 1.80-1.87   & 3360-3650   &\ldots  &\ldots \\
GY194   & 2.19-2.24   & 3550-3865   &\ldots  &\ldots \\
GY253   & 7.08-7.59   & 3350-3820   & 8.2    & 3350  \\
GY410   & 2.51-2.53   & 3870-4120   & 2.7    & 4060  \\
B162730 & 2.44-2.49   & 3555-3850   &\ldots  &\ldots \\
WL19    & 14.46-16.45 & 6000-7210   &\ldots  &\ldots \\

\hline
\end{tabular}

(a) This work; (b) \citet{LR99}
\end{table}

\begin{figure} 
\psfig{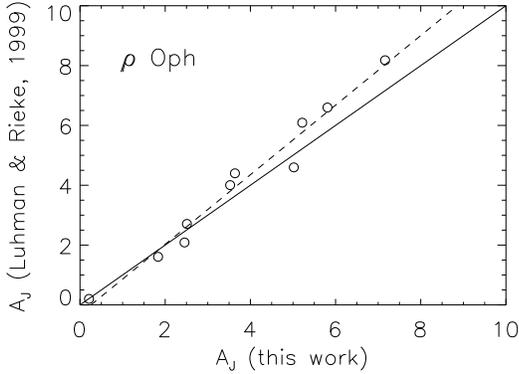}
\caption{Comparison of the values of the dust extinction $A_\mathrm{J}$ derived from spectroscopy \citep[]{LR99} and those derived from the IR SED fitting (this work) for 10 Class~III sources in $\rho$ Oph. The dashed line is the linear fitting line. The solid line is the equality line $A_\mathrm{J}$(this work) = $A_\mathrm{J}$ \citep[]{LR99}.} \label{comparison_Av}
\end{figure}

\subsection{IR SED fitting}\label{SED_fitting}

Different methods may be used to derive the extinction by dust grains. Because the spectra of Class~III sources lie in the near-IR range, the extinction $A_\mathrm{J}$ is directly derived, instead of the more usual $A_\mathrm{V}$. Available optical/IR spectroscopy leads to the spectral type determination and intrinsic colors which, when compared to the observed colors and an extinction law, give an estimate of the dust extinction.  This is the case for most sources in IC\,348 \citep[]{LUH98,LUH99}, Cha~I \citep[]{LAW96,GAU92}, R CrA \citep[]{WAL97}, ONC \citep[]{HIL97} and for some objects in $\rho$ Oph \citep[]{LR99}.  These sources generally have low or moderate extinction, allowing optical/IR spectroscopy.  

If the spectral type is unknown, the dust extinction is determined by fitting the near to mid-IR spectral energy distribution (SED) with a blackbody spectrum.  This is the case of some embedded sources in $\rho$ Oph where $A_\mathrm{J}$ can be as high as 14 magnitudes ($A_\mathrm{V}$ $\simeq$ 50 mag) and optical/near-IR spectroscopy cannot lead to an accurate determination of the spectral type.  We use the $I$ band ({\it DENIS}), $JHK$ photometry ({\it 2MASS}) and 6.7 $\mu$m and 14.3 $\mu$m photometry from ISOCAM \citep[]{BON01} to construct the SED.  For some sources, we also have $L$(3.5$\mu$m), $M$(4.8$\mu$m) and $N$(10$\mu$m) photometry \citep[]{GRE94}.

The wavelength dependence of the interstellar extinction and polarization provides constraints on the characteristics of interstellar grains.  It is well-established that the extinction law depends only on the ratio of total to selective extinction $R_\mathrm{V} = A_\mathrm{V}/E(B-V) \simeq 3.1$ for the diffuse interstellar medium \citep[]{RL85, CAR89} and is higher in some regions. An important point is that, as shown by \citet{CAR89}, {\it the determination of $A_\mathrm{J}$ is independent of $R_\mathrm{V}$}, contrary to the determination of $A_\mathrm{V}$ (see below). 

\citet{DRA89} reviewed the extinction by interstellar dust at IR wavelengths and showed that for wavelengths between 0.7 and 7 $\mu$m the observed extinction, both in the Galaxy and in the Magellanic clouds, is consistent with a power-law 
$A_\mathrm{\lambda}$ $\sim$ $\lambda^{-\alpha}$ with index $\alpha$ $\simeq$ 1.75.  \citet{CAR89} adopt the \citet{RL85} curve with $\alpha$ $\simeq$ 1.6.  \citet{MAR90} established a power law extinction curve with $\alpha$ $\simeq$ 1.8 extending from 0.8 to 4.8 $\mu$m for both diffuse and dense clouds. In this work, we adopt $\alpha$ = 1.8 but consider the effect of values ranging from 1.6 to 2.0.  Beyond 4.8 $\mu m$, we use the extinction law given by \citet{LUT96} based on the ISO/SWS observation of the galactic center.

We fit the SED of each object using the absorbed blackbody model \citep[]{MYE87}. The photospheric flux of the source $F_\mathrm{bb}(T_\mathrm{bb})$ is assumed to be a blackbody at temperature $T_\mathrm{bb}$. 

$$F_\mathrm{obs}=F_\mathrm{bb}(T_\mathrm{bb}) \times e^{-\tau_{\lambda}}$$

with $$\tau_\lambda=\frac{A_\mathrm{J}}{1.086} \times (\frac{A_\lambda}{A_\mathrm{J}}) = \frac{A_\mathrm{J}}{1.086} \times (\frac{\lambda(\mu m)}{1.25})^{-\alpha}$$

\noindent (valid for $\lambda \geq 1 \mu$m), where $F_\mathrm{obs}$ and $F_\mathrm{bb}$ are the observed and blackbody flux in Jansky, $\tau_{\lambda}$ is the optical depth at the wavelength $\lambda$ and $\alpha$ is the power law index. We assume that the dust emission is negligible. The best estimate of $A_\mathrm{J}$ is found by using a least-squares fit to the blackbody model. The fitting parameters are then $T_\mathrm{bb}$, $A_\mathrm{J}$. We vary the power law index $\alpha$ from 1.6 to 2.0 to obtain the resulting uncertainties on the parameters.  

Eleven of 20 X-ray emitting Class III sources in $\rho$ Oph have IR spectra and therefore known spectral types \citep[]{LR99}. These authors derived the absorption $A_\mathrm{J}$ using near IR photometry and spectral type. The nine other Class~III sources of $\rho$ Oph lack IR spectroscopy. We derive a blackbody fit of the SED of these objects to estimate the extinction value, which is given in Table \ref{bbfit}. Fig. \ref{X-ray spectra and SED} illustrates a sample of three fits corresponding to the three X-ray sources analyzed in the previous section.

\begin{figure*}
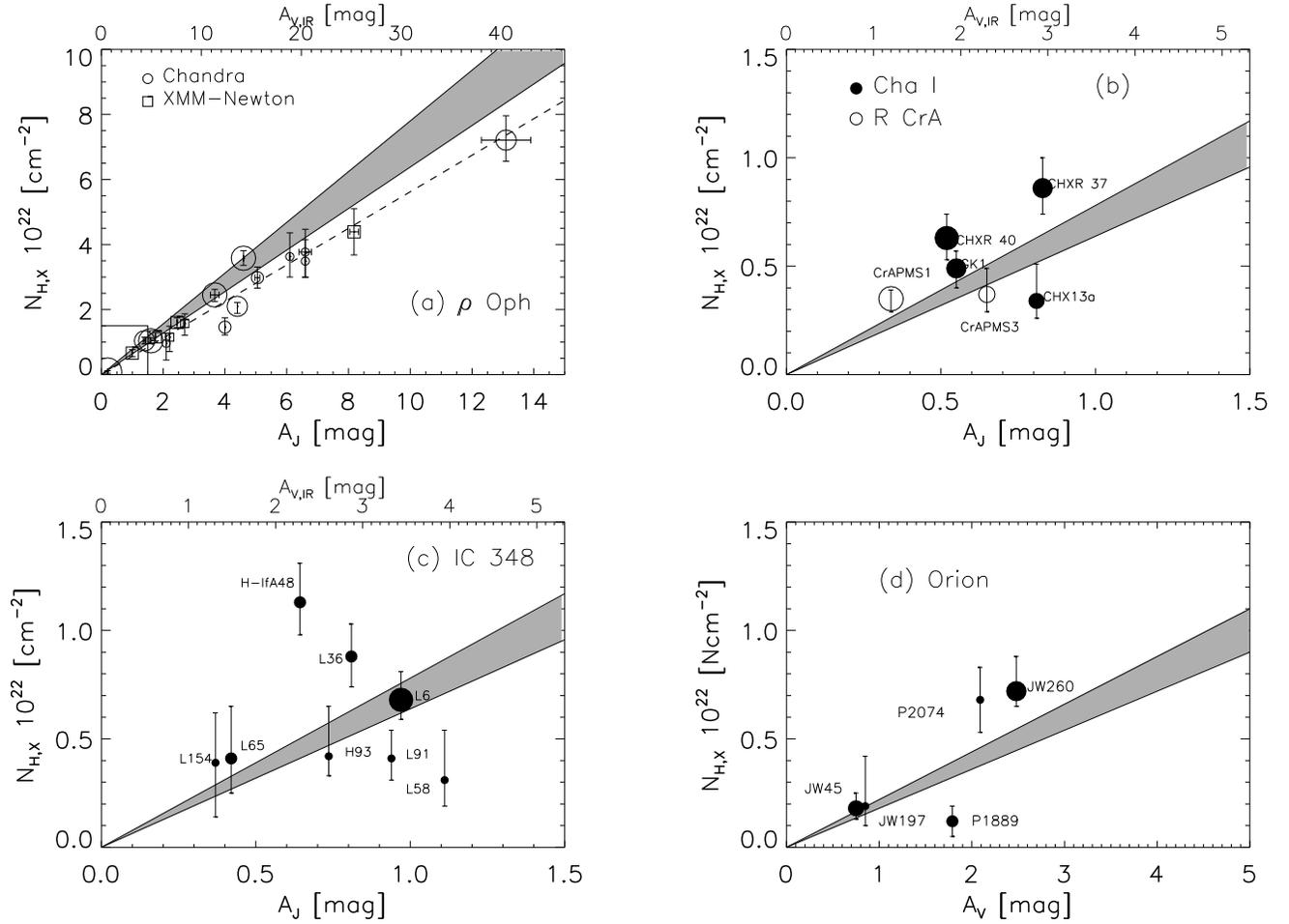
 
\begin{tabular}{cc}
\psfig{file=MS3731.f6a,width=9cm} &
\psfig{file=MS3731.f6b,width=9cm} \\
\psfig{file=MS3731.f6c,width=9cm} &
\psfig{file=MS3731.f6d,width=9cm} \\
\end{tabular}
\caption{X-ray absorption column density $N_\mathrm{H,X}$ versus visual extinction $A_\mathrm{J}$ for Class~III sources in $\rho$ Oph (a), R\,CrA and Cha~I (b), IC\,348 (c), Orion Nebula (d) using the default standard solar ISM abundances in {\it XSPEC}. The error bars in X-ray column densities are represented by vertical bars at 90\% confidence (1.6$\sigma$). The dashed line in $\rho$ Oph is the fitting curve: $N_\mathrm{H}$=5.57($\pm$ 0.35) $\times$ 10$^{21}$ $A_\mathrm{J}$ (cm$^{-2}$). The shaded area is the range of published galactic gas/dust relationship (see Table \ref{NH_AV_values}). The symbol size is proportional to the number of photons detected in X-ray spectra in log (from 195 to 7378). Note that $A_\mathrm{J}$ is up to 14 mag ($A_\mathrm{V}$ $\simeq$ 45 mag) in the $\rho$ Oph region whereas $A_\mathrm{J}$ $<$ 1.5 mag ($A_\mathrm{V}$ $<$ 5 mag) in the other regions.} \label{Nx.Av.all}
\end{figure*}

\subsection{Comparison with other extinction determinations}

Table \ref{bbfit} compares the parameters ($A_\mathrm{J}$, $T_\mathrm{bb}$) derived from the SED fitting with the values from \citet{LR99}. The two values of the extinction and the temperature represent the range resulting from the variation of the power-law index $\alpha$ from 1.6 to 2.0.  \citet{LR99} used $K$-band spectroscopy in $\rho$ Oph to obtain determine spectral types and extinctions. Our fitting values are generally in agreement with those given in the literature when available.

Fig. \ref{comparison_Av} compares the values of the dust extinction $A_\mathrm{J}$ derived from $K$-band spectroscopy \citep[]{LR99} and those derived from the IR SED fitting (This work). A linear relation can be found: $A_\mathrm{J}$ [\citep[]{LR99}] =1.16 $\times$ $A_\mathrm{J}$ [This work] $-$ 1.06. The values of $A_\mathrm{J}$ from spectroscopy is $\sim$ 10\% larger than our SED fitting values. 

\begin{table}
\caption[]{The $N_\mathrm{H}$/$A_\mathrm{V}$ ratios from the 
literature. (See also Table 1.)}\label{NH_AV_values}
\small
\begin{tabular}{cc}
\hline
\hline
$N_\mathrm{H}$/$A_\mathrm{V}$  & Ref.\\
\hline
1.8$\times$10$^{21}$ cm$^{-2}$ (a) & \citet{BOH78} \\
1.9$\times$10$^{21}$ cm$^{-2}$ (a) & \citet{WHI81} \\
2.2$\times$10$^{21}$ cm$^{-2}$ (a,b) & \citet{GOR75, RYT96} \\
2.0$\times$10$^{21}$ cm$^{-2}$ (b) & \citet{RYT75} \\
1.8$\times$10$^{21}$ cm$^{-2}$ (b) & \citet{PRE95}\\
\hline
(a) UV; (b) X-rays. 

\end{tabular}
\end{table}

\section{The $N_\mathrm{H,X}/A_\mathrm{J}$ ratio in nearby star forming regions \label{new_ratio}}

\subsection{Comparison with the galactic value}\label{Nh-Aj-ratio}

Fig. \ref{Nx.Av.all}a shows the correlation between the hydrogen column density derived from the X-ray spectra $N_\mathrm{H,X}$, and the dust extinction $A_\mathrm{J}$ derived from optical/IR spectroscopy for 19 bona-fide Class III TTS in the $\rho$ Oph cloud, assuming the default solar ISM abundances in {\it XSPEC}.  The extinction $A_\mathrm{J}$ is probed up to 14 mag ($A_\mathrm{V}$ $\simeq$ 45 mag). For the sources where $A_\mathrm{J}$ is determined from the blackbody model, the horizontal error bars represent the range in $A_\mathrm{J}$ obtained by varying the index $\alpha$ from 1.6 to 2.0 (\S \ref{SED_fitting}).  The resulting best-fit linear relation is $N_\mathrm{H,X}$=5.57 ($\pm$ 0.35) $\times$10$^{21}$ $A_\mathrm{J}$ (cm$^{-2}$ mag$^{-1}$).

However, the gas-to-dust ratio is usually studied with the help of $N_{\rm H}/A_{\rm V}$. In our Galaxy, this ratio is $(N_\mathrm{H}/A_\mathrm{V})_\mathrm{gal}= 1.8 - 2.2 \times 10^{21}$ cm$^{-2}$ per visual magnitude (Table \ref{NH_AV_values}). This galactic relation, obtained using X-ray absorption, and valid up to $A_\mathrm{V}$ $\simeq$ 30 mag \citep[][see Table 1]{GOR75, RYT75, RYT96, PRE95} is consistent with the one determined using Ly$\alpha$ absorption in the UV limited to low extinctions ($A_\mathrm{V} \simlt$ 2-3 mag).

We want now to compare these values with that obtained above from the $N_{\rm H}/A_{\rm J}$ correlation. This requires the conversion of $A_{\rm V}$ into $A_{\rm J}$, which depends on the shape of the dust extinction curve. \citet{CAR89} have shown that this shape, for a variety of interstellar regions, is well represented with a single parameter $R_{\rm V}=A_{\rm V}/(A_{\rm B}-A_{\rm V})$ where $A_{\rm B}$ is the dust extinction in the B-band. From this parameterization, they found that $A_{\rm J}/A_{\rm V}=0.4008-0.1187(R_{\rm V}/3.1)^{-1}$. We use this empirical relationship, and the average galactic value of $R_{\rm V}=3.1$, to convert $A_{\rm V}$ into $A_{\rm J}$, yielding the galactic gas-to-dust correlation in the $J-$band: $(N_{\rm H}/A_{\rm J})_\mathrm{gal}=6.4-7.8 \times 10^{21}$ cm$^{-2}$ mag$^{-1}$. This range is indicated by the shaded area in Fig. \ref{Nx.Av.all}, and compared with our results for $\rho$ Oph: it is clear from Fig. \ref{Nx.Av.all} that {\it the $N_{\rm H}/A_{\rm J}$ correlation lies significantly below ($\simgt 2\sigma$) the galactic values}.

The four other regions (Cha~I, R\,CrA, IC\,348 and Orion) are probed only to low extinctions, $A_\mathrm{J}$ $\simlt$ 1.5 mag ($A_\mathrm{V}$ $<$ 4).  Fig. \ref{Nx.Av.all}b compares the hydrogen column density and the dust extinction toward 4 bona-fide Class~III sources in the Cha~I dark cloud.  Three points deviate from the galactic relation.  Fig. \ref{Nx.Av.all}c and \ref{Nx.Av.all}d represent 8 lines of sight toward IC\,348 and 5 lines of sight toward the Orion Nebula Cluster (ONC).  Note that Fig.  \ref{Nx.Av.all}d displays directly $A_\mathrm{V}$, instead of $A_\mathrm{J}$: the values of $A_\mathrm{V}$ for Class~III sources in ONC are taken from \citet{HIL97} using optical spectroscopy. Due to the small sample in each of these regions, we cannot obtain a relationship between $N_\mathrm{H,X}$ and $A_\mathrm{J}$ like in $\rho$ Oph. We note that the comparison of $N_\mathrm{H,X}$ and $A_\mathrm{V}$ has already been done by \citet{FEI02} for the ONC and \citet{GET02} for NGC\,1333, but only in global terms, for instance without distinguishing between Class~III sources and Class~II sources (TTS with disks), or with automated fits. In these studies, the general behavior shows a large dispersion, and no correlation between $N_{\rm H}$ and $A_\mathrm{J}$ or $A_\mathrm{V}$ can be found. For comparison, there are only two data points in $\rho$ Oph having $A_\mathrm{J}$ $<$ 1.5, so that spanning a large range in $A_\mathrm{J}$ is crucial to look for a correlation.

In conclusion, for $\rho$ Oph we find that (i) a tight correlation exists between $N_{\rm H}$ and $A_{\rm J}$ up to $A_{\rm J} \sim 15$, and (ii) this correlation lies significantly below the galactic correlation obtained by converting $A_\mathrm{V}$ into $A_\mathrm{J}$ using the canonical value $R_\mathrm{V} = 3.1$ and $N_\mathrm{H}/A_\mathrm{V}$ determined from UV and X-ray measurements. For the four other regions studied (Cha~I, R~CrA, IC~348, and the ONC), the data probe a too small extinction range ($A_{\rm J} \leq 1.5$), and no similar correlation can be found.

\subsection{Spatial dependence of the $N_\mathrm{H}/A_\mathrm{J}$ ratio}

\subsubsection{High extinction: $\rho$ Oph}\label{high_ext}

\begin{figure*} 
\psfig{file=MS3731.f7,width=12cm} 
\caption{Spatial distribution of our sample of Class~III sources in $\rho$ Oph. Background: DSS $R$ optical image. Foreground: {\it ISOCAM} 7$+$15 $\mu m$ image \citep[]{ABE96}. This combination background map is taken from \citet{GRO00}. The DCO$^+$ map of \citet{LOR90} is shown by contours and the dense cores are labeled. The circles ($\circ$) are sources with $N_\mathrm{H,X}/A_\mathrm{J}$ normalized by the regression line (5.57 $\times$ $10^{21}$) with a correlation $\pm$ 10\%. The triangles up ($\bigtriangleup$) are sources with normalized values of $N_\mathrm{H,X}/A_\mathrm{J}$ higher than 1.1 and the triangles down ($\bigtriangledown$) are lower than 0.9. The dashed lines are {\it Chandra} and {\it XMM} field-of-views.} \label{carte_oph} 
\end{figure*}

We now investigate whether the deviation from the galactic $N_{\rm H}$ vs. $A_{\rm J}$ correlation found in $\rho$ Oph, and the dispersion of the $N_{\rm H}$ vs. $A_{\rm J}$ or $A_{\rm V}$ data points, depend on the location of the sources within the clouds. Fig.  \ref{carte_oph} symbolizes $N_\mathrm{H,X}/A_\mathrm{J}$ along individual lines of sight in relation with optical, mid$-$IR and molecular maps. Lines-of-sight with values close (within 10\%) to the $N_{\rm H}$ vs. $A_{\rm J}$ correlation are indicated by circles; those for which the $N_{\rm H}/A_{\rm J}$ ratio deviates by more than $+$ or $-$ 10\% are indicated by triangles, pointing upwards or downwards respectively. We find no spatial correlation between the value of the observed $N_\mathrm{H,X}/A_\mathrm{J}$ ratio and the DCO$^{+}$ contours indicating the dense cores. 

We now discuss more precisely the two sources with the highest $A_\mathrm{J}$, particularly WL\,5 which plays an important role in the $N_\mathrm{H,X}$--$A_\mathrm{J}$ correlation:  WL5 ($A_\mathrm{J} \simeq$ 14) and WL19 ($A_\mathrm{J} \simeq$ 16).

$\bullet$ WL 5:  This source is located in the core B and has been extensively studied at many wavelengths: radio by \citet{AND87}, mm observation by \citet{AND94}, near and mid-IR by \citet{GRE94}. \citet{GRE95} determined an F7 spectral type, 
bolometric luminosity 33.5 L$_{\sun}$ and extinction $A_\mathrm{V}$=51 mag, using near-IR spectroscopic observations. 
This value of $A_\mathrm{V}$ is overestimated because \citet{GRE95} used the conversion $A_\mathrm{J}/A_\mathrm{V}=0.265$, which corresponds to $R_\mathrm{V} = 2.7$ \citep[]{CAR89}. The 100\,ks {\it Chandra} observation provided 1322 photons for this source. WL5 is also detected in the {\it XMM-Newton} observation but its spectrum is contaminated by a Class~II source (IRS~37) located only 7$\arcsec$ away. Given the large number counts in the {\it Chandra} spectrum,  there is no particular problem, and $N_\mathrm{H.X}$ is accurately derived (Fig. \ref{X-ray spectra and SED}). This source is included in Fig. \ref{Nx.Av.all}a.

\begin{figure*}
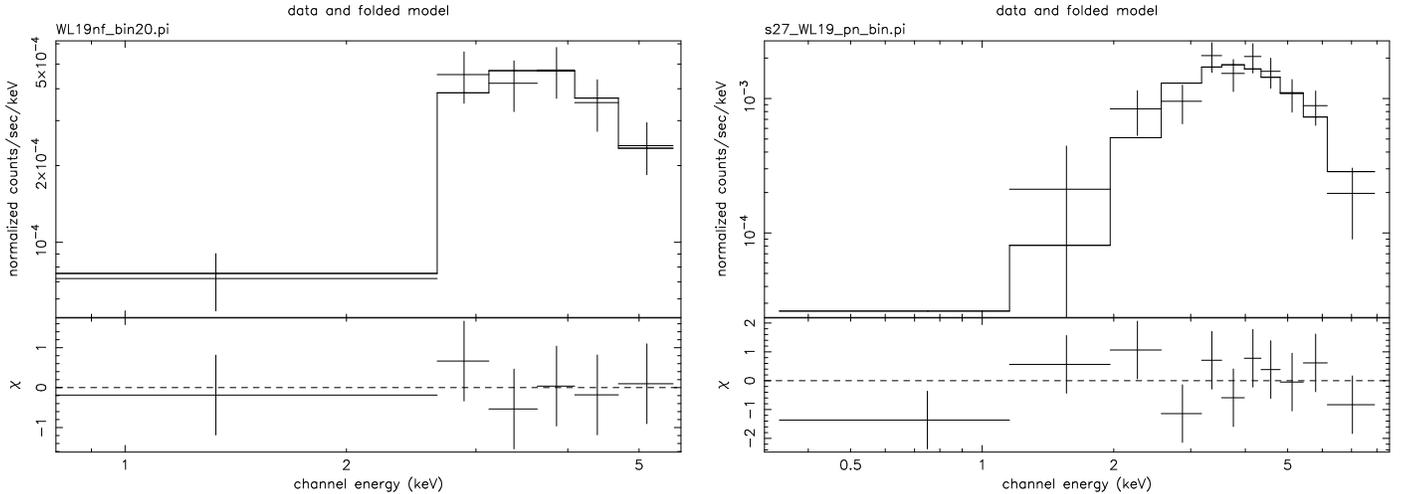
 
\begin{tabular}{cc}
\psfig{file=MS3731.f8a,width=9cm,angle=-90} &
\psfig{file=MS3731.f8b,width=9cm,angle=-90} \\
\end{tabular}
\caption{{\it Chandra} (left) and {\it XMM-Newton} (right) spectra of WL19 ($N_\mathrm{H,X}=7-14 \times 10^{22}$ cm$^{-2}$, $T_\mathrm{X}=1.5-3$ keV, $L_\mathrm{X} = 0.4-1.5 \times 10^{30}$ erg.s$^{-1}$, using the default standard solar ISM abundances in {\it XSPEC}).}\label{WL19}
\end{figure*}

$\bullet$ WL 19:  This object was classified as a Class I YSO by \citet{WIL89} and as a Class II by \citet{AND94}.  \citet{BON01} detected no mid-IR excess from this source and concluded that it is likely a luminous Class III source located {\it behind} the cloud (i.e. not embedded within the cloud) with very high extinction.  This source was observed by both {\it Chandra} and {\it XMM-Newton}.  The 100\,ks {\it Chandra} spectrum provides only about 120 photon counts and the 30\,ks XMM spectrum has 150 counts.  The spectra are shown in Fig. \ref{WL19}. The values of $N_\mathrm{H,X}$ derived from these two spectra with low statistics are $\sim$ 2 $\sigma$ apart. For this reason, we have not included this source in the $N_\mathrm{H,X}/A_\mathrm{J}$ relation.


\subsubsection{Low extinctions}

Unlike $\rho$ Oph, the other star forming regions have only one {\it Chandra} observation and their spatial coverage is limited by the ACIS field-of-view (17$\arcmin \times17\arcmin$).  We could build only a small sample of Class~III sources in Cha~I, R~CrA, IC~348, and the ONC. In these regions, Fig. \ref{Nx.Av.all}b, c, d show large deviations of the $N_\mathrm{H,X}/A_\mathrm{J}$ ratio ($> \pm 10 \%$) from the galactic value. Fig. \ref{dss_IC348} shows the spatial distribution of our sample of Class~III sources in IC~348. The background image is the DSS $I$ band image, and the contours represent {\it IRAS} 60 $\mu m$ isophotes, taken as column density tracers. As in $\rho$ Oph, we use circles and triangles pointing upwards and downwards to symbolize deviations of the $N_{\rm H}/A_{\rm J}$ ratio, here with respect to the galactic value. Again, no correlation with column density is apparent. The reasons from the deviations observed in Fig. \ref{Nx.Av.all}c are unclear, although they would probably play a minor role if the available range in $A_\mathrm{J}$ was larger as in $\rho$ Oph.

\begin{figure} 
\psfig{file=MS3731.f9,width=9cm} 
\caption{DSS $I$ band image and spatial distribution of our sample Class~III sources in IC\,348 with {\it IRAS} 60 $\mu m$ contours overlaid. The circles ($\circ$) represent sources with $N_\mathrm{H,X}/A_\mathrm{J}$ normalized by the galactic value of $N_\mathrm{H}/A_\mathrm{J}$ with a correlation $\pm$ 10\%. The triangles up ($\bigtriangleup$) are sources with normalized values of $N_\mathrm{H,X}/A_\mathrm{J}$ higher than 1.1 and the triangles down ($\bigtriangledown$) are lower than 0.9. } \label{dss_IC348} 
\end{figure}

\section{Constraints on the gas and dust properties from the $N_\mathrm{H,X}/A_\mathrm{J}$ ratio}\label{discussion}

The properties of the gas and dust in $\rho$ Oph can be constrained from our measurement of the gas-to-dust ratio. Indeed, our value of the slope of the $N_{\rm H,X}-A_{\rm J}$ correlation in $\rho$ Oph is lower than in the Galaxy. This may arise {\em (i)} because we underestimate the total gas column density $N_{\rm H,X}$ (through the metallicity $Z$) or, {\em (ii)} because the grain properties (abundance with respect to the gas, extinction per unit mass in the $J-$band) are modified toward the $\rho$ Oph cloud.

\subsection{Effect of metallicity}

\begin{figure*}
\begin{tabular}{cc} 
\psfig{file=MS3731.f10a,width=9cm} &
\psfig{file=MS3731.f10b,width=9cm} \\
\psfig{file=MS3731.f10c,width=9cm} &
\psfig{file=MS3731.f10d,width=9cm} \\
\end{tabular}
\caption{Comparison between the derived $N_\mathrm{H,X}$ using ``new'' solar abundances \citep[]{HOL01, ALL01, ALL02} and (a) those using ``old'' solar abondances, (b) those using photospheric F \& G star abundances \citep[]{SOF01a, RED03}, (c) those using diffuse ISM abundances \citep[]{WIL00} and (d) those using diffuse ISM B-star abundances \citep[]{SOF01a}. The solid line is the equality line.}\label{comparison_abund} 
\end{figure*}

As recalled in \S \ref{introduction}, the total hydrogen column densities are inferred from the photoelectric cut-off observed in the X-rays. Such a cut-off depends on the column density of metals $N_\mathrm{Z}$, which is then converted into hydrogen column density $N_\mathrm{H}$ by assuming a set of element abundances, irrespective of whether the heavy elements, mostly C, N, O (which we will refer to as ``metallicity'') are in the gas or in the grains.

Solar abundances (determined from analysis of the solar photosphere or meteorites) have generally been used as the reference abundance for the ISM \citep[]{AND82, GRE89, GRE98}, and this is what we have used so far in this paper: as noted previously, these abundances are the default in {\it XSPEC}. We now refer to these abundances as ``old'' solar abundances. Recently, \citet{HOL01, ALL01, ALL02} published a ``new'' determination of photospheric solar abundances, in particular for carbon and oxygen. Therefore, we have included these ``new'' solar abundances in {\it XSPEC} and re-calculated the derived $N_\mathrm{H,X}$ from our X-ray spectral fits. We find that for a given $A_\mathrm{J}$, $N_\mathrm{H,X}$ is $\simeq$ 20\% higher than when using the ``old'' solar abundances. This is illustrated in Fig. \ref{comparison_abund}a, where we compare the X-ray derived hydrogen column densities using the ``old'' solar and the ``new'' solar abundances. We find: $N_\mathrm{H}$ (new~solar) = 1.32 $\times$ $N_\mathrm{H}$ (old~solar) $-$ 0.15 $\times$ 10$^{22}$ cm$^{-2}$. The linear relation between $N_\mathrm{H,X}$ and $A_\mathrm{J}$ for $\rho$ Oph becomes: $N_\mathrm{H}$(new~solar)~=~7.24($\pm$ 0.46)~$\times$~10$^{21}$~$A_\mathrm{J}$~(cm$^{-2}$).
This new relation is then in remarkable agreement with the relation $(N_{\rm H}/A_{\rm J})_\mathrm{gal}=6.4-7.8 \times 10^{21}$ cm$^{-2}$ mag$^{-1}$ quoted above for the Galaxy when using the galactic value $R_\mathrm{V} = 3.1$.

Abundance measurements have also been made in the local diffuse ISM with the $HST$, toward field B stars \citep[]{WIL00, SOF01a}. More recently, \citet{RED03} presented photospheric abundances of 181 F \& G disk stars. We shall refer to these abundances as the "diffuse ISM Wilms et al. 2000", "diffuse ISM B star" and "photospheric F \& G star" abundances, respectively. These total gas plus dust abundances are 20-30\% lower than the ``old'' solar abundances (depending on the element) \citep[]{SNO96, CAR96, MEY97, SOF01a}. 
Therefore, we have re-run our X-ray spectral fits for the $\rho$ Oph sources using these abundances. Figs. \ref{comparison_abund}b, c, d show that the derived X-ray column densities using these diffuse ISM and photospheric abundances are almost equal ($\pm$~$\sim$~10\%). Noting that these recent measurements probe relatively short lines of sight ($\simlt$ 1 kpc for B stars and $\simlt$ 150 pc for F \& G stars). We shall define this new set of abundances simply as the ``local'' abundances.

We conclude that the lower value of $N_{\rm H,X}/A_{\rm J}$ for $\rho$ Oph may be fully accounted for by adopting the metallicity corresponding to the ``local'' abundances.

\subsection{Effect of grain properties}\label{Rv_observations}

However, recall that several works \citep[\eg][]{VRB93} showed that in $\rho$ Oph, $R_\mathrm{V}$ is not galactic. In this section, we keep the ``old'' solar abundances and discuss the effect of grain properties in relation with $R_\mathrm{V}$.

As shown in Appendix A, the $N_\mathrm{H,X}/A_\mathrm{J}$ ratio is proportional to the gas-to-dust mass ratio $M_\mathrm{gas}/M_\mathrm{dust}$ (hereinafter G/D). Thus, the low $N_\mathrm{H,X}/A_\mathrm{J}$ ratio could be explained if G/D = 70 (instead of the average galactic value of 100). Assuming that this higher dust mass is due to the formation of ice mantles \citep[]{TAN90}, about 80~\% of the gas phase oxygen should be depleted onto grains (with [O/H]$_{\rm gas}$ = 3 10$^{-4}$, \citet{SAV96}). We do not consider this possibility in greater detail however because the depletion in $\rho$ Oph is not expected to be strong (see Appendix B). 

\vspace{0.3cm}
Another possibility to lower the $N_{\rm H,X}/A_{\rm J}$ ratio is that the $A_{\rm J}/A_{\rm V}$ ratio is larger. This ratio and the shape of the extinction curve for $\lambda \leq$ 1 $\mu$m depend on $R_{\rm V}$. This parameter is closely related to the mean size of dust grains as shown in Fig. A.1. Changes of $R_{\rm V}$ thus trace processes affecting the dust size distribution. In particular, $R_{\rm V}$ is expected to increase in cold, dense molecular clouds \citep[]{CAR89} where grain coagulation probably occurs \citep[]{STE03}. 

From polarimetry in optical bands, \citet{VRB93} found $R_{\rm V}$ = 4 in a peripheral region of the $\rho$ Oph cloud where $A_{\rm V}\leq $ 3 mag. To extend this study to regions with larger $A_{\rm V}$'s requires near-IR or mid-IR polarimetry, which has never been performed before. To derive $R_{\rm V}$ in $\rho$ Oph, we assume that $N_{\rm H,X}/A_{\rm V}$ is constant throughout the Galaxy, equal to 1.9$\times$10$^{21}$ cm$^2$ mag$^{-1}$ \citep[$\pm 20 \%$, \eg][]{KIM96}. We thus find $R_{\rm V}$  = 6.0 $\pm$ 2.5, \ie $A_{\rm J}/A_{\rm V}=$ 0.34$^{+0.02}_{-0.04}$ (instead of 0.28 for $R_{\rm V}$ = 3.1) using the parameterization of the extinction curve given by \citet{CAR89}. We note that this value brackets that of \citet{VRB93}. Note that because of the functional dependence between $R_{\rm V}$ and $A_{\rm J}/A_{\rm V}$, the uncertainty on $R_{\rm V}$ is large. Therefore, with the present measurement of $N_{\rm H,X}/A_{\rm J}$, it is difficult to reach a firm conclusion on the value of $R_{\rm V}$ and on its possible variation from the periphery to the center of the cloud. 

\subsection{Constraints on the dust size distribution in $\rho$ Oph}

We now use a physical model of gas and dust properties to constrain the grain size distribution and the gas-to-dust mass ratio in $\rho$ Oph. The details of the grain modeling are described in Appendix A.

Using the observationnally derived value of $R_\mathrm{V}$ for $\rho$ Oph, including the uncertainties (see \S \ref{Rv_observations}), we find the minimum grain size $a_\mathrm{min}=$ 0.025 $\mu$m for $R_\mathrm{V}=4$ and 0.07 $\mu$m for $R_\mathrm{V}=6$ (see Fig. \ref{Av.Nx.model.oph}). These values are $\sim$ 5$-$10 times larger than in the galactic diffuse medium. Fig. \ref{Av.Nx.model.oph} shows that the average grain size $<a>$ increases with $a_\mathrm{min}$. For $\rho$ Oph, it is $\sim$ 0.035$-$0.095 $\mu$m instead of 8$\times$10$^{-3}$ $\mu$m (Eq. (A.4)) in the galactic diffuse medium. We also find that the corresponding gas-to-dust mass ratio G/D for $\rho$ Oph lies between 80 and 95 using the MRN size distribution as described in Appendix A. 
This is to be compared with \citet{SAV96} who investigated G/D using $HST$ abundance measurements toward $\zeta$ Oph. When calibrating this measurement with solar abundances, they found G/D $\simeq$ 72. This number corresponds to {\it diffuse} medium at {\it low} extinction ($A_\mathrm{V} \simeq$ 1.2 mag). Our derived values of G/D for $\rho$ Oph is consistent with this value within $\sim$ 10\%. We have however used other dust grain models, like \citet{DES90}, which use a different grain composition and slightly different extinction cross-sections, and found G/D $\sim 130$. While this value seems to be ruled out by the work of \citet{SAV96}, it shows that the conclusion is rather sensitive to the grain composition. Also,  the inspection of Fig. \ref{Av.Nx.model.oph} shows that a wide range of values of grain size parameters ($a_\mathrm{min}$, $a_\mathrm{max}$, index $q$) is compatible with a large value of $R_\mathrm{V}$. 

\subsection{Relationship between metallicity and $R_{\rm V}$}

\begin{figure} 
\psfig{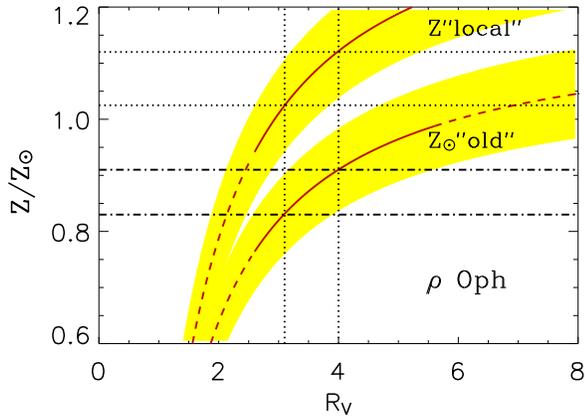}
\caption{Metallicity $Z$ as a function of $R_\mathrm{V}$ for $\rho$ Oph using the ``old'' solar and ``local'' abundances. The shaded area represents the uncertainties. The dashed line is the extension of the empirical relation of \citet{CAR89} valid for $2.6 \leq R_\mathrm{V} \leq 5.6$. The dotted and dashed-dotted lines indicate the metallicity for $R_\mathrm{V}=$ 3.1 and 4 \citep[]{VRB93} using the ``local'' and ``old'' solar abundances.}\label{Rv-Z} 
\end{figure}

Both explanations (``local'' metallicity or larger $R_{\rm V}$) for a lower $N_{\rm H,X}/A_{\rm J}$ ratio in $\rho$ Oph are plausible and the values of metallicity and $R_{\rm V}$ quoted above correspond to extreme cases. Either grain coagulation has occurred and $R_\mathrm{V} \sim 6$ (\S \ref{Rv_observations}), or the standard galactic $R_\mathrm{V}=3.1$ applies but the galactic $N_\mathrm{H}/A_\mathrm{V}$ relation was derived from X-ray measurements using the ``old'' solar abundances. The ``local'' metallicity is indeed lower than the ``old'' solar values as shown in Figs \ref{comparison_abund}. Therefore, we have looked for an independent indication in $\rho$ Oph from the CO/H$_2$ ratio derived from millimeter-wave measurements. But as explained in Appendix B, CO may be depleted onto grains, and above all the uncertainties on this ratio are much larger than the $\sim$ 20$-$30\% accuracy needed to test metallicity values.

The trade-off between these two interpretations of our results -- $R_\mathrm{V}$ vs. $Z$ --  can be expressed in a quantitative manner. We now combine them to show they compensate each other: this will be helpful when more observational constraints on the metallicity and/or $R_{\rm V}$ come about. Using the \citet{CAR89} empirical relation between $R_\mathrm{V}$ and $A_\mathrm{J}/A_\mathrm{V}$ (found for $R_\mathrm{V}$ = 2.6 to 5.6), we obtain a relation between the cloud metallicity relative to solar, $Z/Z_\mathrm{\odot}$, and $R_\mathrm{V}$~:

\begin{eqnarray}
\frac{Z}{Z_\mathrm{\odot}} = \frac{(N_\mathrm{H}/A_\mathrm{J})_\mathrm{\rho\,Oph}}{(N_\mathrm{H}/A_\mathrm{V})_\mathrm{gal}} \times \Big[0.4008 - \frac{0.3679}{R_\mathrm{V}}\Big]
\end{eqnarray}

\noindent taking $N_{\rm H}/A_{\rm V} = 1.9\times \, 10^{21}$ cm$^2$ mag$^{-1}$ throughout the $\rho$ Oph cloud, and using the fact that the hydrogen column density derived from the X-ray spectra is almost inversely proportional to the metallicity $Z$ for $0.4 \simlt Z/Z_\odot \simlt 1.2$. Beyond this range, the higher order terms of the $N_\mathrm{H,X}-Z$ relation can no longer be neglected.

This Z($R_\mathrm{V}$) relation is plotted in Fig. \ref{Rv-Z} using the ``old'' solar and ``local'' abundances.  The solid portion of the curve represents the most likely region where the $\rho$ Oph dense gas lies, constrained by $Z \sim Z_\mathrm{local}$ and $R_\mathrm{V}=3.1$.  The shaded area shows the uncertainties on $R_{\rm V}$ (resulting from the uncertainties on $N_\mathrm{H,X}$/$A_\mathrm{J}$, $\sim 10$\%) and on $Z/Z_\odot$ (from Eq.(1)).  
As discussed qualitatively above, the $\rho$ Oph metallicity is close to the ``new'' solar abundances for $R_\mathrm{V}$ close to 3.1 when assuming that the galactic gas-to-dust ratio remains unchanged. This is consistent with the local ISM abundances toward B stars and photospheric F \& G star abundances.

\section{Conclusions and implications}\label{conclusion}

We have used the X-ray spectra of 41 young stars without circumstellar matter (Class~III IR sources, or equivalently ``weak-line'' T Tauri stars) obtained with {\it Chandra} and {\it XMM-Newton}, to derive the hydrogen column densities $N_\mathrm{H,X}$ in six nearby molecular clouds, and we have determined or re-determined the corresponding dust extinction for these stars in the $J-$band, $A_\mathrm{J}$. Only in the $\rho$ Oph cloud do we have a large enough sample (19 stars) to establish a linear correlation between $N_\mathrm{H,X}$ and $A_\mathrm{J}$:
\medskip

$N_\mathrm{H,X}/A_\mathrm{J} (Z_\mathrm{\odot old})=5.6 (\pm 0.4) \times 10^{21} ({\rm cm^{-2} mag^{-1}}),$

$N_\mathrm{H,X}/A_\mathrm{J} (Z_\mathrm{local})=7.2 (\pm 0.5) \times 10^{21} ({\rm cm^{-2} mag^{-1}}),$

\medskip
\noindent using the ``old'' solar and ``local'' abundances for the X-ray extinction, respectively. These relations are valid up to $A_\mathrm{J} \simlt 14$ ($A_\mathrm{V} \simlt$ 50), \ie probe for the first time the gas-to-dust ratio in the dense interstellar medium up to very large extinctions. Also, we find that there is no significant deviation of the $N_\mathrm{H,X}$/$A_\mathrm{J}$ ratio along the various lines of sight toward the X-ray sources, \ie no significant difference as a function of their respective column densities.

Comparing with the similar relation for the Galaxy (Table 5), $(N_\mathrm{H}$/$A_\mathrm{J})_\mathrm{gal}$= 6.4$-$7.8 $\times$ 10$^{21}$ cm$^{-2}$ mag$^{-1}$, obtained by using the usual value of the total-to-selective extinction ratio $R_\mathrm{V} = 3.1$ (which gives $A_\mathrm{J}/A_\mathrm{V} = 0.28$), we find that the $\rho$ Oph $N_\mathrm{H,X}/A_\mathrm{J} (Z_\mathrm{\odot old})$ relation lies significantly ($\simgt 2\sigma$) below the galactic relation.

We show that this result is consistent with the recent downwards revision of the solar abundances. We find that the lower value of $(N_\mathrm{H}$/$A_\mathrm{J})_\mathrm{\rho Oph}$ previously determined using the ``old'' solar abundances changes by 20\% when using the new set of abundances. But the galactic relation derived from the X-ray absorption (see Table \ref{NH_AV_values}) was obtained using ``old'' solar abundances. Therefore, the difference between the $\rho$ Oph and Galactic relation persists and can be accounted for entirely by a difference in metallicity.

In order to compare with the galactic relation derived from previous X-ray absorption measurements, we keep the ``old'' abundances. We investigated the possibility that this was due to a change in $R_\mathrm{V}$, resulting from a change in dust grain properties (size distribution, composition, etc.). We find $R_\mathrm{V} = 6.0 \pm 2.5$.

Quantitatively, the combination of the dark cloud metallicity $Z/Z_\odot$ and of $R_\mathrm{V}$ (Fig. \ref{Rv-Z}) shows that the metallicity in $\rho$ Oph is close to the ``local'' abundances (the Sun, photospheric F \& G stars and diffuse ISM toward B stars) for $R_\mathrm{V}$ close to 3.1 if the galactic gas-to-dust ratio remains unchanged.

By contrast, a similar study on five other star-forming regions observed in X-rays turned out to have  a limited value, with only a few X-ray sources with sufficiently high count statistics, and insufficiently well-defined IR properties. These studies do not go deeper than $A_\mathrm{J} \sim 1.5$ ($A_\mathrm{V} \sim 5$), preventing us to derive a reliable $N_\mathrm{H,X}$/$A_\mathrm{J}$ ratio.

The use of the updated solar metallicity and the relative difference between the {\it local} ($d \simlt$ 1kpc) and the {\it galactic ISM at large distances} (up to 4 kpc) gas-to-dust extinction relation has some interesting implications:

\begin{enumerate}

\item A first implication is on the chemical evolution of the Galaxy. By cross-correlating X-ray absorption and abundance measurements, we have shown that the metallicities of the Sun, of nearby F \& G stars, of the local diffuse ISM and dense ISM (in the case of the $\rho$~Oph cloud) are all identical within $\sim$~10\%. This seems {\it a priori} paradoxical, since the associated timescales for the nucleosynthetic evolution of the Galaxy are very different, being respectively 5 Gyr, $\sim 1$ Gyr, a few Myr, and $\sim 1$ Myr: we would expect the youngest objects (namely the dense ISM) to be chemically the most evolved, i.e., to have larger metallicities. This ``lazy nucleosynthesis'' is in contrast with the strong chemical enrichment observed over kpc scales towards the center of the Galaxy. \citet{MIS02} have noted the almost flat behavior of the metallicity as a function of galactic radius in the solar vicinity (which is now confirmed by our measurement of the metallicity of the $\rho$~Oph cloud). According to the chemical diffusion model proposed by these authors, this can be explained by a slow mixing and low star formation rate at the galactic solar radius, resulting from to the fact that the solar radius lies close to the galactic corotation radius.

\item A second (related) implication is on the metallicity of the Galaxy. We have shown using X-rays that the metallicities of the $\rho$~Oph cloud (using the local abundances) and of the Galaxy (also using X-rays, but with the `old'' solar abundances) differ by $\simgt 20$\%. This can be interpreted as the Galaxy being ``overmetallic'' with respect to the $\rho$~Oph cloud and the local medium (stars + ISM). As recalled above, the Galaxy does show an enhanced metallicity towards the galactic center. This means that (i) the X-ray absorption measurements are indeed quite sensitive to the ISM metallicity provided the absorption column density is high enough, and (ii) that by a numerical coincidence, the $\sim 20$\% downwards revision of the local abundances matches the metallicity difference between the local medium and the galaxy {\it averaged over the long lines of sight of the galactic X-ray absorption measurements} (see Tables \ref {NH_AV_ratio} and \ref{NH_AV_values}). A detailed modeling of the metallicity along the directions towards distant galactic X-ray sources (compact binaries and SNR, up to $\sim 4$ kpc), using \citet{MIS02} for example, would be required to demonstrate this quantitatively, but this is beyond the scope of the present paper.

\item A third implication is on grains in dense clouds. The main conclusion is that no strong deviations from the usual grain properties are found, like size or composition, even in dense regions, as constrained by the behavior of $R_\mathrm{V}$ using grain models. However, the uncertainties are large, and some qualitatively important changes (like coagulation in the densest regions, CO depletion on grains, etc.) could go unnoticed. But if one could independently and precisely measure the cloud metallicity (for instance by measuring the photospheric metallicity of the young stars they gave birth to), {\it one could estimate $R_\mathrm{V}$ in dense regions} from the curve in Fig. \ref{Rv-Z}. This may then better constrain the grain properties. As far as we are aware, there has been up to now only one attempt at measuring the metallicity of young low-mass stars by \citet{PAD96}. She determined the iron abundances of about 30 PMS stars in nearby clouds, and found their [Fe/H] to be solar within $\pm 0.15$ dex: this uncertainty however is larger than the precision level of $\sim 10$\% which we need on the metallicity, so more accurate measurements are needed.

\item A fourth implication is more practical, and has to do with the determination of X-ray luminosities. In recent years, as illustrated in this paper, X-ray spectra from the new generation of X-ray satellites have allowed to directly measure hydrogen column densities, which in turn allow to correct the X-ray luminosities for extinction. In all cases of the existing literature, {\it XSPEC} fits were made by default using ``old'' solar ISM abundances. Our work shows that {\it XSPEC} should use abundances revised downwards by $\sim 20$\% for distances up to $\simlt$ 1kpc, with the resulting (upwards) corrections on $L_\mathrm{X}$ for sources in nearby clouds.

\end{enumerate}

The present work should be considered as a pilot study, illustrating various successes and problems which can be attacked using X-ray absorption in molecular clouds in relation with IR extinction data.  More work, on more clouds, including new X-ray and IR observations, for which existing archival data are clearly not optimized or insufficient, is needed to extract all the potentialities of the method.

\begin{acknowledgements}

We would like to thank Malcolm Wamsley, Charlie Ryter, M. G\"udel, A. Jones and S. Madden for their helpful discussion. We are grateful to the referee J. Mathis for his comments. 
The DENIS project is supported in France by the Institut National des Sciences de l'Univers, the Education Ministry and the Centre National de la Recherche  Scientifique, in Germany by the State of Baden-W\"urtemberg, in Spain by the DGICYT, in Italy by the Consiglio Nazionale delle Ricerche, in Austria by the Fonds zur F\"orderung der wissenschaftlichen Forschung und Bundesministerium f\"ur Wissenschaft und Forschung.
This publication makes use of data products from the Two Micron All Sky Survey, which is a joint project of the University of Massachusetts and the Infrared Processing and Analysis Center/California Institute of Technology, funded by the National Aeronautics and Space Administration and the National Science Foundation.

\end{acknowledgements}

\appendix

\section{Modeling the interstellar grain population}\label{modeling}

\subsection{Modeling}

We model $N_\mathrm{H}$/$A_\mathrm{J}$ and $R_\mathrm{V}$ using the \citet{DRA84} extinction cross section and the MRN grain size distribution \citep[]{MAT77}.

$N_\mathrm{H}$/$A_\mathrm{J}$ depends on the gas-to-dust mass ratio $M_\mathrm{gas}$/$M_\mathrm{dust}$ (hereinafter G/D) and the extinction cross section $Q_\mathrm{J,ext}(a)$ averaged over a size distribution of grains $n(a)$, where $a$ is the grain (assumed spherical) radius. Absorption and scattering cross sections are calculated for grains of radius up to 1 $\mu m$ \citep[]{DRA84}. We consider a mixture of silicate (53\%) and graphite (47\%) as described by \citet{DRA84}. The grain size distribution is described by a power law: $ n(a)da \propto a^{-q} da,$ $a_\mathrm{min} \leq a \leq a_\mathrm{max}$, and $n(a)da$ is the density of grains along the line of sight with radii in the range $a$ to $a+da$ and $q=$3.5, $a_\mathrm{min}$=0.005 $\mu m$ and $a_\mathrm{max}$=0.25 $\mu m$ \citep{MAT77}. Specifically,

\begin{eqnarray}
\frac{N_\mathrm{H}}{A_\mathrm{J}} = \frac{M_\mathrm{gas}}{M_\mathrm{dust}} \times \frac{1}{m_\mathrm{H}} \times \frac{M_\mathrm{dust}}{\int_{a_\mathrm{min}}^{a_\mathrm{max}} Q_\mathrm{J,ext}(a) \pi a^2 n(a) da}
\end{eqnarray}

\begin{eqnarray}
\noindent {\rm with}\;\; M_\mathrm{dust}=\int_{a_\mathrm{min}}^{a_\mathrm{max}} \rho(a) \frac{4}{3} \pi a^3 n(a) da.
\end{eqnarray}

The dust mass density $\rho(a)$ is assumed to be 3.3 g~cm$^{-3}$ for silicate and 2.26 g~cm$^{-3}$ for graphite.

$R_\mathrm{V}$ only depends on the extinction cross section at V (0.55 $\mu m$) and B (0.44 $\mu m$) bands respectively, averaged over the grain size distribution:

\begin{eqnarray}
R_\mathrm{V} = \frac{<a^2 Q_\mathrm{V,ext}(a)>}{<a^2 Q_\mathrm{B,ext}(a)> - <a^2 Q_\mathrm{V,ext}(a)>}.
\end{eqnarray}

We note that {\em (i)} $R_\mathrm{V}$ does not depend on the ratio G/D and {\em (ii)} $R_\mathrm{V}$ is very sensitive to the grain composition (graphite or silicate) and to the dust size distribution. The simplest way to obtain $R_\mathrm{V}$ is to measure it in $B$ and $V$ bands but this is very difficult for sources with $A_\mathrm{V} \simgt 4-5$ mag. One can then use sub-mm observations to constrain the size distribution and/or optical properties.

The mean grain size is defined as follows:

\begin{eqnarray}
<a> = \frac{\int_{a_\mathrm{min}}^{a_\mathrm{max}} a \, n(a) da}{\int_{a_\mathrm{min}}^{a_\mathrm{max}} n(a) da}
\end{eqnarray}

\begin{figure}
\psfig{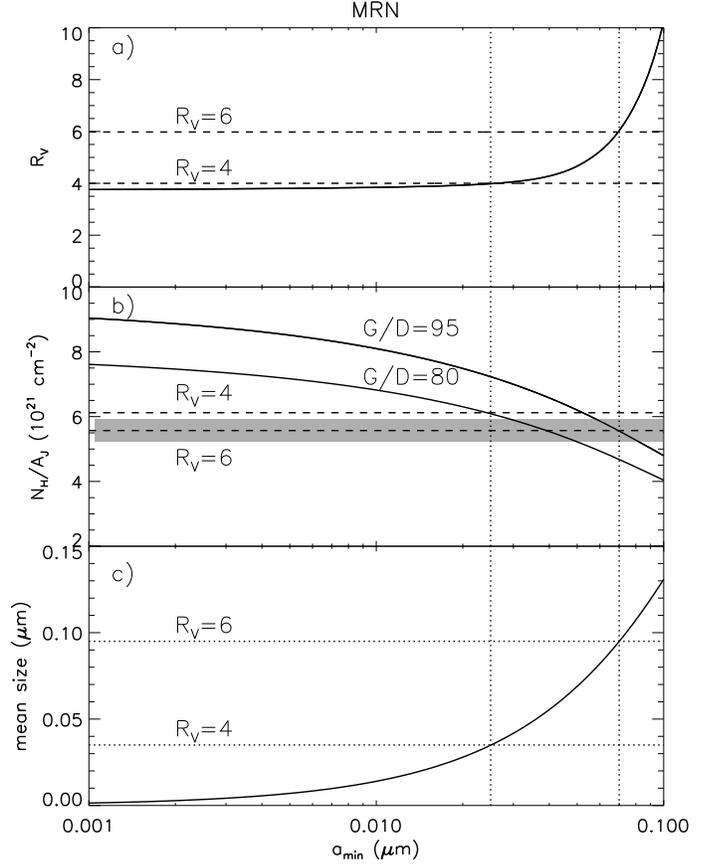}
\caption{$R_\mathrm{V}$ (a), gas-to-dust extinction ratio $N_\mathrm{H}/A_\mathrm{J}$ (b), and the mean grain size (c) as a function of $a_\mathrm{min}$. The dashed line and the shaded area in (b) are the mean value $\pm 1\sigma$ of the observed $N_\mathrm{H,X}$/$A_\mathrm{J}$ in $\rho$ Oph. The dotted lines represent the interval of $a_\mathrm{min}$ for $R_\mathrm{V}$ between 4 and 6.} \label{Av.Nx.model.oph}
\end{figure}

The solid curves in Fig. \ref{Av.Nx.model.oph}a, b, c show $R_\mathrm{V}$ (a), the gas-to-dust extinction ratio $N_\mathrm{H}/A_\mathrm{J}$ (b) and the mean grain size (c) as a function of $a_\mathrm{min}$ as derived from Eqs (A.3), (A.1) and (A.4), respectively. The two dashed lines in Fig. \ref{Av.Nx.model.oph}b represent the values of $N_\mathrm{H,X}/A_\mathrm{J}$ for $R_\mathrm{V}=6$ (or $Z/Z_\odot=1$) and $R_\mathrm{V}=4$ (or $Z/Z_\odot=0.9$) (see Fig. \ref{Rv-Z}). For $R_\mathrm{V}$ between 4 and 6, we find that G/D is comprised between 95 and 80 and $a_\mathrm{min}$ between 0.025 and 0.07 $\mu$m.

\vspace{-0.3cm}

\section{The $N$(CO)/$N$(H$_2$) ratio and CO depletion onto grains}\label{CO}

We discuss here whether we can test the conclusion from our combined X-ray/IR analysis that the ISM abundance of heavy atoms (or metallicity) is lower than solar in $\rho$ Oph, using the CO/H$_2$ ratio derived from X-ray and millimeter observations.
We take as a test case WL~5, the most deeply extincted source in our sample ($A_\mathrm{V} \sim 45$; see \S 4.2.1), using the total hydrogen (mostly H$_2$) column density $N_\mathrm{H,X}$ derived from the X-ray spectrum. \citet{WIL83} have done C$^{18}$O observations of the dense cores of the $\rho$ Oph cloud. They determined the column density of C$^{18}$O adopting a local thermodynamic equilibrium (LTE) approach. Along the line of sight toward WL\,5, the C$^{18}$O column density is 1.53$\times$10$^{16}$ mag.cm$^{-2}$ \citep[]{WIL83}.  Since H$_2$ is the dominant component of H along the line of sight, the H$_2$ column density is N(H$_2$)=$N_\mathrm{H,X}$/2. To convert N(C$^{18}$O) to N($^{13}$CO), we use the isotopic abundance ratios [$^{12}$C/$^{13}$C]=77 and [$^{16}$O/$^{18}$O]=560 \citep[]{WIL94}.  
For WL\,5, we find N(H$_2$)/N($^{13}$CO)$\sim$3.3 $\times$10$^5$ using solar abundances, and N(H$_2$)/N($^{13}$CO)$\sim$4.5 $\times$10$^5$ using B-star abundances. These values are consistent with the usual relationship N(H$_2$)/N($^{13}$CO)=5.0$\pm$2.5 $\times$ 10$^5$ by \citet{DIC78}. Note that the determination of N(CO) does not depend on metallicity. In conclusion, given the errors we are not sensitive to changes in the CO/H$_2$ ratio, whether we take solar or B-star abundances. Since we are looking for evidence of underabundances, the fact that CO can be depleted on grains in dense clouds (see \citet{ALV99} for a study of the correlation between C$^{18}$O line emission and dust extinction toward the L\,977 molecular cloud) is an additional difficulty. We can only exclude an underabundance of CO (intrinsic or by depletion) by a factor $>$ 2, i.e., much larger than the effect we are looking for.

\normalsize

\end{document}